\documentclass[a4paper,fleqn]{cas-dc}

\usepackage[authoryear]{natbib}
\usepackage{caption}
\usepackage{balance}
\usepackage{stfloats}

\def\tsc#1{\csdef{#1}{\textsc{\lowercase{#1}}\xspace}}
\tsc{WGM}
\tsc{QE}
\tsc{EP}
\tsc{PMS}
\tsc{BEC}
\tsc{DE}

\newcommand{\todo}[1]{}
\renewcommand{\todo}[1]{{\color{red} TODO: {#1}}}
\newcommand{\commentout}[1]{}

\begin{document}
\let\WriteBookmarks\relax
\def\floatpagepagefraction{1}
\def\textpagefraction{.001}
\shorttitle{Capstone courses in software engineering}

\shortauthors{Saara Tenhunen et~al.}

\title [mode = title]{A systematic literature review of capstone courses in software engineering}                      

\author{Saara Tenhunen*}[type=editor,
                        auid=000,bioid=1,
                        orcid=0000-0002-4894-8365]




\ead{saaraten@gmail.com}

\credit{Conceptualisation of this study, Methodology, Validation, Formal analysis, Investigation, Data curation, Writing - original Draft, Writing - Review \& Editing, Visualisation}

\address{University of Helsinki, , , Finland}

\author{Tomi Männistö*}
\ead{tomi.mannisto@helsinki.fi}
\credit{Conceptualisation of this study, Methodology, Writing - original Draft, Writing - Review \& Editing, Supervision}

\author{Matti Luukkainen}
\ead{matti.luukkainen@helsinki.fi}
\credit{Conceptualisation of this study, Methodology, Writing - original Draft, Writing - Review \& Editing, Supervision}

\author{Petri Ihantola}
\ead{petri.ihantola@helsinki.fi}
\credit{Conceptualisation of this study, Methodology, Writing - original Draft, Writing - Review \& Editing, Supervision}

\cortext[cor1]{Corresponding author}

\begin{abstract}
Context: Tertiary education institutions aim to prepare their computer science and software engineering students for working life. While much of the technical principles are covered in lower-level courses, team-based capstone projects are a common way to provide students with hands-on experience and teach soft skills.\newline 
Objective: This paper explores the characteristics of software engineering capstone courses presented in the literature. The goal of this work is to understand the pros and cons of different approaches by synthesising the various aspects of software engineering capstone courses and related experiences.\newline 
Method: In a systematic literature review for 2007--2007, we identified 127 primary studies. 
These studies were analysed based on their presented course characteristics and the reported course outcomes. \newline 
Results: The characteristics were synthesised into a taxonomy consisting of duration, team sizes, client and project sources, project implementation, and student assessment. We found out that capstone courses generally last one semester and divide students into groups of 4--5 where they work on a project for a client. For a slight majority of courses, the clients are external to the course staff and students are often expected to produce a proof-of-concept level software product as the main end deliverable. The courses also offer versatile assessments for students throughout the project. \newline 
Conclusions: This paper provides researchers and educators with a classification of characteristics of software engineering capstone courses based on previous research. We also further synthesise insights on the reported outcomes of capstone courses. Our review study aims to help educators to identify various ways of organising capstones and effectively plan and deliver their own capstone courses. The characterisation also helps researchers to conduct further studies on software engineering capstones. \newline
\end{abstract}

\begin{highlights}
\item Our taxonomy of course features is based on ACM/IEEE guide for capstone courses
\item There is a vast diversity in how capstone courses in SE are implemented
\item More research is needed to compare different course implementation strategies  
\item Many of the capstone courses are shorter than the recommended two semesters
\item Many of the capstone courses are missing an external client
\end{highlights}


\maketitle

\section{Introduction}

Universities and other tertiary education institutions should provide their students with sufficient skills and abilities before the students enter working life. In software engineering-related programs, this entails having an understanding of the common principles and theory in computer science \citep{ieeeCS, ieeeSE} and technical competencies and knowledge demanded by the industry \citep{radermacher2014investigating, garousi2019closing}. Any recent graduate should also be able to apply this technical knowledge in practice \citep{ieeeCS}.

While much of the technical knowledge and theories are covered in lower-level courses, many institutions hold team-based capstone project courses to ensure students are ready to apply the knowledge in a workplace environment. A ``capstone course`` usually means a course that finishes an academic degree \citep{ikonen2009discovering}. The main goal of a capstone project is to provide hands-on experience in applying the tools, techniques, principles and best practices that are taught more theoretically in previous courses \citep{ziv2010capstone, majanoja2018reflections, panicker2020exposing}. Capstone projects are also regarded as crucial in teaching students the necessary soft skills such as teamwork \citep{keogh2007scalable, venson2016academy}, verbal and written communication \citep{watkins2010competitive}, time management \citep{dupuis2010experiments}, problem solving \citep{majanoja2018reflections} and project management \citep{haddad2013one}. In computer science (CS) and software engineering (SE) programs, capstone courses generally last one or two semesters, and they include assigning students into teams and having them work on various kinds of software engineering projects \citep{ikonen2009discovering, bowring2016shaping, paasivaara2019collaborating}. In these projects, they are expected to experience stages of the software development life-cycle from requirements solicitation to software maintenance \citep{keogh2007scalable}. 

Given the general acceptance of capstones as a practical way of teaching industry-relevant skills, a high number of institutions have implemented their own capstone courses. This has resulted in a great deal of research done on capstone courses and their outcomes. In order to provide a coherent and compact view of software engineering capstones, this research synthesises the current body of knowledge on the topic in a systematic manner. We believe that such a review gives educators an effective tool for planning and implementing their own capstone courses. Researchers can also benefit from a systematic review of capstones to conduct further comparative studies on the impact of the varying course forms.

This study is organised as follows. The next section focus on the previous literature reviews on SE capstones, as well as general characteristics of such courses. Section \ref{research_method} describes the research questions and the related methods, including how the articles were selected. Section \ref{results} presents the results of the literature review. The main findings and their validity are discussed in Section \ref{discussion}. Suggestions for future research are also given. Finally, Section \ref{conclusions} concludes the research.

\section{Previous work}\label{previous_work}

\subsection{Systematic literature reviews of SE capstones}\label{systematic_literature_reviews}

Many literature reviews have been written in the general area of software engineering education (SEE). Usually, they focus on specific sub-areas of SEE such as teaching methods in software engineering \citep{anicic2022teaching}, practical approaches to SEE \citep{marques2014systematic}, trends in SEE \citep{cico2019industry, cico2021exploring} or teaching global software engineering \citep{fortaleza2012towards}. 

As the focus of this research is especially on project-based capstone courses in software engineering, we carefully sought any earlier systematic reviews done on them. The search was conducted on May 17th, 2022, first in the citation database Scopus and secondly in Google Scholar. Table \ref{tab:review_search_terms} lists the search terms used in both databases. All search results produced by Scopus were checked to see whether they include an SLR of SE capstones, whereas for Google Scholar, each search produced tens of thousands of hits, so we went through the first 20 pages of each search (200 hits). At this point, the results started to become highly irrelevant, and often repetitive. Based on our search, we believe that the three review papers presented in Table \ref{tab:capstone_reviews} are the ones that have been published so far on this topic. Table \ref{tab:capstone_reviews} also presents the characteristics of capstone courses each of these reviews investigated. Next, we will briefly present these studies and discuss the necessity of this review.

\begin{table*}[width=1\textwidth,cols=3]
\scriptsize
\caption{Searches for systematic reviews on software engineering capstones}\label{tab:review_search_terms}
\begin{tabular}{p{0.1\linewidth}p{0.77\linewidth}p{0.05\linewidth}}
\toprule
Database & Search term & Hits \\
\midrule
Scopus & TITLE-ABS-KEY( software  AND  engineering AND capstone AND literature AND review ) & 10 \\ 
Scopus & TITLE-ABS-KEY( software AND engineering AND capstone AND review ) & 61 \\
Scopus & TITLE-ABS-KEY( education  AND  ``software engineering``  AND  literature  AND  mapping ) & 40 \\
Scopus & TITLE-ABS-KEY( project  AND  course  AND  software  AND  engineering  AND  systematic  AND  review ) & 20 \\
Scopus & TITLE-ABS-KEY( ``computer science`` AND capstone AND literature AND review ) & 8 \\
Scopus & TITLE-ABS-KEY( software  AND engineering  AND project  AND course AND systematic  AND literature  AND review ) & 17 \\
Scopus & TITLE-ABS-KEY( ``software engineering`` AND education AND literature AND review ) & 182 \\

Google Scholar & software engineering capstone systematic review & 20 300 \\
Google Scholar & software engineering capstone characteristics & 31 400 \\
Google Scholar & computer science capstone literature review & 53 400 \\
Google Scholar & capstone literature review & 94 800 \\

\bottomrule
\end{tabular}
\end{table*}

\begin{table*}[width=1\textwidth,cols=4,pos=h]
\footnotesize
\caption{Systematic reviews of software engineering capstones}\label{tab:capstone_reviews}
\begin{tabular}{p{0.3\linewidth}p{0.05\linewidth}p{0.17\linewidth}p{0.38\linewidth}}
\toprule
Title & Year & Ref & Course characteristics examined in the survey \\
\midrule
A survey of computer science capstone course literature & 2011 & \citep{dugan2011survey} & Course-related: models, learning theories, goals, topics, student evaluation, evaluation.\newline Project-related: software process models, phases, type, documentation, tools, groups, instructor administration. \\
Designing the IT capstone course & 2019 & \citep{martin2019designing} & Course duration, learning of new skills, project identification and selection, teams sizes, team formation, followed methodologies, assessment of learning outcomes, team and project supervision* \\
A review of literature on assessment practices in capstone engineering design courses: Implications for formative assessment & 2006 & \citep{trevisan2006review} & Connection to student achievement \\
\bottomrule
\end{tabular}
*For the survey by \cite{martin2019designing}, these are characteristics, which would have been examined in the actual survey.
\end{table*}

\cite{dugan2011survey} presents a survey done on the literature related to undergraduate computer science capstone courses. The survey is comprehensive, comprising 200 papers on the subject and summarising them under two major themes: course issues and project issues (Table \ref{tab:capstone_reviews}). Out of these, course issues include aspects related to the general course organisation, such as course models, learning theories present in the course and student evaluation. Project issues, on the other hand, categorise and describe the projects and how they are implemented. The category includes things like software process phases, project type and documentation of the projects.

\cite{martin2019designing} has provided an abstract of a systematic literature review for designing an IT capstone course. The review plans to provide answers to several questions relating to capstone course design, such as the optimal team size for project teams, identifying and selecting suitable projects and determining the correct duration for the course (Table \ref{tab:capstone_reviews}). We are not aware that the research  proposed in the abstract would have been completed. 

\cite{trevisan2006review} have performed a systematic review on the assessment practices in capstone engineering design courses. They were especially interested in discovering the extent to which classroom assessment has received attention in the capstone literature. The paper included 32 journal articles and conference proceedings presenting varying assessment techniques and their use.

In addition to the presented three literature reviews, a study on the dimensions of SE and CS capstone projects has been conducted by \cite{burge2009dimensions}. Said dimensions are roughly divided into two groups: project dimensions such as customer identity and development dimensions such as project type and source code visibility. The purpose of their study is to provide a framework for analysing capstone courses, especially in terms of risk and realism. While their categorisation is versatile, their study does not, however, include a thorough systematic literature review. The categorisation presented is more of an experience-based proposal, and therefore the study is left out of Table \ref{tab:capstone_reviews}. 

To the best of our knowledge, there is no extensive, recent literature review done on software engineering capstone courses. A survey conducted by \cite{dugan2011survey} is comprehensive but dated to 2011 and therefore does not cover the large number of primary studies published in the past decade. It also does not provide any quantitative statistics of the course characteristics, which would enable educators or researchers to assess how common some aspect in reality is. \cite{trevisan2006review} provide a review on capstone literature, but it is limited to continuous assessment techniques and is dated to 2006. \cite{martin2019designing} aims to provide a systematic literature review on IT capstone design and characteristics, but as of now, the paper has not proceeded beyond the original abstract. In light of this, current research does not provide an up-to-date view of how SE capstone courses generally are organised and with what kind of outcomes. Such a view on the software engineering capstones would not only provide educators with an important tool for planning their own capstone courses but also give researchers a basis for performing comparative studies on these courses.

\subsection{Background: Capstone course characteristics}\label{capstone_characteristics}

ACM/IEEE Curriculum Guidelines for Software Engineering (SE) Degree Programs \citep{ieeeSE} view the capstone project as an essential element of a SE degree programme and state that the main goal of a capstone course is to ensure that the curriculum has a significant real-world basis. According to \cite{ieeeSE}, incorporating real-world elements into the curriculum is necessary to enable effective learning of software engineering skills and concepts. The ACM/IEEE Curriculum Guidelines for Computer Science (CS) degree programs \citep{ieeeCS} align with these views and state that all graduates of CS programs should have been involved in at least one substantial project. Such projects should challenge students by being integrative, requiring evaluation of potential solutions and working on a larger scale than typical course projects. For students, a capstone project typically represents a culmination of their studies and is one of the last milestones before graduation \citep{ieeeCS, ieeeSEGrad}. Indeed, since the 1970s, hundreds of primary studies have been written on this large, final-year project course \citep{dugan2011survey}.

The \cite{ieeeSE} also lists a set of key recommendations that a capstone course should follow. The recommendations are listed word by word in Table \ref{tab:capstone_recommendations}. We decided to use these recommendations as the basis for formulating our research questions. They give a general outline of capstone courses and therefore provide a valid starting point for the categorisation done in this research.
\begin{table}[width=1\linewidth,cols=2,pos=h]
\footnotesize
\renewcommand{\arraystretch}{1.3}
\caption{ACM/IEEE recommendations for SE capstones}\label{tab:capstone_recommendations}
\begin{tabular}{p{0.1\linewidth}p{0.8\linewidth}}
\toprule
CR \# & Recommendation  \\
\midrule
\hypertarget{CR1}{CR 1} & The project should span a full academic year, giving students adequate time to
reflect upon experiences and retry solutions as appropriate.\\
\hypertarget{CR2}{CR 2} & Where possible, this should preferably be undertaken as a group project. If such
factors as assessment make this difficult, it is essential that there should be a
separate group project of substantial size. \\
\hypertarget{CR3}{CR 3} & Where possible, a project should have a “customer” other than the supervisor so that the student gains fuller experience with product development life-cycle activities. \\
\hypertarget{CR4}{CR 4} & A project should have some form of implementation as its end deliverable so that
the students can experience a wide set of software development activities and
adequately evaluate these experiences. Theory-based projects such as the
development of formal specifications is therefore inappropriate for this role. \\
\hypertarget{CR5}{CR 5} & Evaluation of project outcomes should go beyond concept implementation (“we
built it, and it worked” \citep{glass2004analysis}), using walkthroughs, interviews, or
simple experiments to assess the effectiveness and limitations of the deliverables. \\
\hypertarget{CR6}{CR 6} & Assessment of a capstone project should consider how effectively software engineering
practices and processes have been employed, including the quality of student reflection
on the experience, and not be based only on the delivery of a working system. \\
\bottomrule
\end{tabular}
\end{table}

Thus according to these guidelines, there are some basic characteristics that capstone courses have. They can be characterised as long and substantial projects (CR1, CR2) that should preferably be completed in a team (CR2). Projects should have customers (CR3) for whom the students are expected to deliver some form of real implementation at the end of the course (CR4). Students should therefore engage in real software development activities and not just complete simple, theory-based assignments provided by the teacher (CR4). Evaluation of the project outcomes should focus not only on the fact that the project ``works``, but also assess the deliverables on how well they have been completed (CR5). Finally, the focus of the course and its assessment should be on software engineering practices and processes and students should give adequate opportunities to reflect on the experience (CR6). The next section describes in more detail the process of how we derived the research questions based on these basic characteristics.

\section{Research questions and method}\label{research_method}

\subsection{Research questions}

Characteristics of capstone courses (described in Section~\ref{capstone_characteristics}) can be achieved in many ways. The main goal of this research was to understand these differences in how capstone courses are implemented in universities and other tertiary education institutions, and thus provide a holistic view over the various capstone course implementations.

We decided to use the ACM/IEEE Curriculum Guidelines for Undergraduate SE Degree Programmes \citep{ieeeSE} as the basis for starting to explore these characteristics. The recommendations are listed in Table \ref{tab:capstone_recommendations}. Although some of these aspects, e.g., team formation, have been addressed in the previous reviews, previous literature reviews are slightly outdated. Moreover, we are not aware of any study covering all the aspects mentioned in the ACM/IEEE recommendations.

Related to \hyperlink{CR1}{CR1}, we were interested in the duration of the courses and what rationale primary studies provide for choosing a specific course duration if any:

\begin{description}
\item[RQ1]{ What is the duration of SE capstone courses, and what advantages or disadvantages are related to a certain duration?}
\end{description}

Related to \hyperlink{CR2}{CR2}, we wanted to find out if these projects are conducted in teams, how teams are composed and what is the rationale behind choosing a certain team size:

\begin{description}
\item[RQ2]{What team sizes do SE capstone courses have, and how are team sizes justified?} 
\end{description}

Based on the ACM/IEEE recommendations (i.e., \hyperlink{CR3}{CR3}), a
project should have a customer other than the teacher of the course. An alternative approach to bringing an outside view to a project is to outsource project topics. Thus, our third research question, \emph{how are the project and client sourcing handled in SE capstone courses}, was divided into two sub-questions:

\begin{description}
\item[RQ3.1] Who acts as the client for capstone projects?
\item[RQ3.2] How are the ideas for projects sourced?
\end{description}

Related to \hyperlink{CR4}{CR4}, we asked: \emph{How are the projects in capstone courses implemented (RQ4)}. We wanted to uncover what students do in these courses and therefore, we looked into the actual project implementation. As `project implementation` can mean a multitude of things, we decided to divide this research question into smaller, more concrete sub-questions:

\begin{description}
\item[RQ4.1] What artefacts are students expected to produce on capstone courses?
\item[RQ4.2] What is the software life-cycle gone through during these projects?
\item[RQ4.3] How are the implementation technologies chosen for capstone projects?
\end{description}

With RQ4.1 we aimed to find out what students actually produce in these courses and whether any software is being developed. RQ4.2 helped us to find out if the capstone project is as integrative experience on software engineering practices as curriculum guidelines \citep{ieeeSE, ieeeSEGrad} suggest. Finally, finding out how educators make the choices for implementation technologies and what implications these choices have, gave some insight into project implementation.

As \hyperlink{CR6}{CR6} speaks about assessment, our last research question also asked: \emph{How is the student assessment conducted on SE capstone courses? (RQ5)}. As assessment can be divided into continuous feedback and final grading, RQ5 was also split into two:

\begin{description}
\item[RQ5.1] How are the students assessed at the end of SE capstone courses?
\item[RQ5.2]  How are students guided, if at all, during SE capstone courses?
\end{description}

The rationalisation here was that we wanted to uncover whether the evaluation is based on a multitude of factors like \citep{ieeeSE} suggests and whether students are given adequate possibilities to reflect on their experiences \citep{hattie2007power}.

In order to get a comprehensive representation of how project-based capstone courses are generally organised, relevant research articles were searched. One could argue that the characteristics and any organisational details of these courses could be derived from the web pages of universities and other tertiary institutions. However, we wanted not only to produce a list of characteristics such as the duration and workload of the courses but also to reveal more about the contents of these courses. An important part of the research was also to provide educators with insights related to the various characteristics. Without any evaluation or assessment of the chosen structure and characteristics, this would have been impossible to achieve. 

\subsection{Search strategy}

The method used in this study follows the SLR method by \cite{kitchenham2007guidelines}. The initial data collection was done by finding relevant sources from scientific databases: Scopus, ACM Digital Library, IEEE Xplore and ScienceDirect. Some preliminary searches were conducted on these databases to find out to which extent research articles use the word ``capstone`` and its synonyms when describing large, degree-culminating project courses in software engineering-related programs. It turned out that the term ``capstone`` is well-known and widely used in research articles. It was also used by \cite{dugan2011survey} in their earlier work. Therefore, the first search string was simply constructed as:
\noindent
\begin{verbatim}
software AND capstone
\end{verbatim}

In order to have a complete picture of the project course landscape in software engineering, a second search was performed using the second search string: 

\noindent
\begin{verbatim}
software  AND 'project course'
\end{verbatim}

This was deemed necessary as not all sources had the word ``capstone`` present in the metadata even though they clearly were describing courses relevant to this research. Searches with the two search strings were conducted sequentially in each database.

\cite{dugan2011survey} used ``software engineering course`` as another search term in their study, but we did not want to limit ourselves to the SE discipline, as relevant software-related courses might be presented, for instance, in computer science. Using only the words ``software`` and ``course`` on the other hand, provided too many irrelevant hits. Scopus alone produced nearly 30 000 hits of which only a small fraction would have been relevant to our study.

\begin{figure}
    \centering
    \includegraphics[width=1\linewidth]{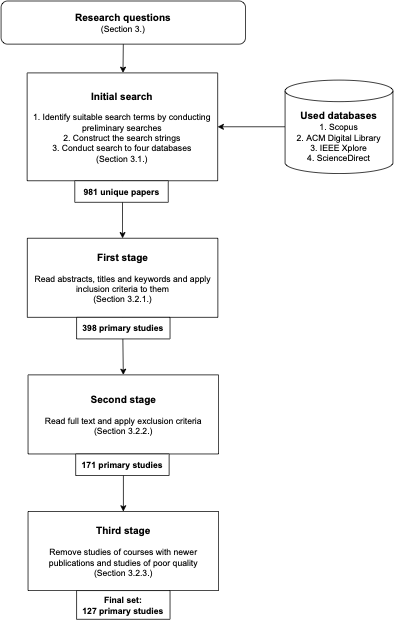}
    \caption{Search strategy}
    \label{fig:search_strategy}
\end{figure}

A total of 981 unique papers were found after combining the papers found from all four databases using the search strings and removing duplicates. The databases were searched on June 11th and June 12th 2022, one after the other, starting with Scopus, moving on to ACM Digital Library, followed by ScienceDirect and finishing with IEEE Xplore. As the search fields and filters are slightly different in each of these databases, the search strings were adjusted to match each specific set-up. They were, however, kept semantically the same across the searches. Exact search strings and initial search results are listed in Table \ref{tab:initial_search_results}. As we wanted to identify current ways of organising capstone courses, the searches in all four databases were limited to the years 2007 to the search day in June 2022. This time period was regarded as sufficiently long to provide a holistic view of the current capstone courses. It overlaps with \cite{dugan2011survey} by a few years but also uncovers 11 years of research done on the area that has not been systematically reviewed since. Three stages of selection were applied to this initial set, after which 127 primary studies remained. Fig. \ref{fig:search_strategy} summarises the search and selection process, and the following subsections will describe it in greater detail.

\begin{table*}[width=1\textwidth,cols=3,pos=h]
\scriptsize
\renewcommand{\arraystretch}{1.2}
\caption{Initial search results}\label{tab:initial_search_results}
\begin{tabular}{p{0.18\linewidth}p{0.66\linewidth}p{0.08\linewidth}}
\toprule
Database & Search strings & Hits  \\
\midrule

Scopus & TITLE-ABS-KEY ( software  AND capstone ) & 762 \\
Scopus & TITLE-ABS-KEY ( software AND ``project course`` ) & 262 \\

ACM Digital Library & [Title: software] AND [Title: capstone]  & 24 \\
ACM Digital Library & [Keywords: software] AND [Keywords: capstone] & 32 \\
ACM Digital Library & [Abstract: software] AND [Abstract: capstone] & 130 \\
ACM Digital Library & [Title: software] AND [Title: ``project course``] & 6 \\
ACM Digital Library & [Keywords: software] AND [Keywords: ``project course``] & 7\\
ACM Digital Library & [Abstract: software] AND [Abstract: ``project course``] & 44 \\

ScienceDirect & (TITLE ABS KEY: SOFTWARE CAPSTONE) & 22 \\
ScienceDirect & (TITLE ABS KEY: SOFTWARE PROJECT COURSE) & 12 \\

IEEE Xplore & (``All Metadata``:Software) AND (``All Metadata``:capstone) & 223 \\
IEEE Xplore & (``All Metadata``:software) AND (``All Metadata``:``project course``) & 86 \\

\bottomrule
\end{tabular}
\end{table*}

\subsection{Paper selection}\label{paper_selection}

The paper selection was conducted from the initial set of 981 sources by the first author (Fig. \ref{fig:search_strategy}). The details of inclusion and exclusion are explained next.

\subsubsection{The first stage - Inclusion criteria}\label{first_stage}

\emph{The titles, abstracts and keywords} of the initial papers were read and evaluated against the inclusion criteria presented below (IC1-IC3). After the first stage, 398 papers remained.
\newline

\noindent
IC1 The title or abstract strongly hints that the study presents frameworks or case studies of software engineering capstones or other large, project-based courses in software engineering \newline
IC2 Based on the title or abstract, the study describes real experiences of implementing a software engineering capstone course \newline
IC3 The title or abstract indicates that the study assesses the outcomes of the course or its characteristics \newline

The first inclusion criterion was developed to set the focus on software engineering courses in particular. A large number of the articles in the initial set were ruled out due to the first criterion (IC1). The papers were found to research, for instance, mechanical engineering courses, which were out of the scope of this research. We also wanted to rule out any purely hypothetical papers, where the researchers show no course that follows the frameworks or structures presented. The second inclusion criterion (IC2) aimed to ensure that all included papers would present a real-world course. The final inclusion criterion (IC3) was generated so that all papers would also evaluate the outcomes of the various course implementations. 

\subsubsection{The second stage - Exclusion criteria}\label{second_stage}

The second stage was performed on the 398 papers remaining from the first stage. Any article that, based on reading the \emph{full paper}, met at least one of the presented exclusion criteria was excluded at this stage. After this selection, 171 articles remained for the final evaluation. The used exclusion criteria were:
\newline

\noindent
EC1 The length of the study is less than four pages 
\newline
EC2 The study is not published in conference proceedings or as a journal article 
\newline
EC3 The study does not have full text available in English
\newline
EC4 The study turned out not to describe a software engineering capstone course in a tertiary institution
\newline
EC5 The study is not able to provide answers to most of the research questions 
\newline

Exclusion criteria from EC1 through EC3 aimed to ensure that the study was of sufficient quality. According to \cite{kitchenham2007guidelines} workshop proceedings often do not provide sufficient input for the purposes of an SLR. Additionally, quite many of the papers that were first published as short workshop proceedings or abstracts were also found to have a conference proceeding or a journal article published later on. EC3 relates to the language skills of the authors as well as the status of English as the primary language in software engineering-related research. These exclusion criteria led to some papers being rejected before reading their entire content.

For exclusion criteria EC4--EC5, the content of the article was examined more carefully and, in most cases, read in its entirety to make a justified decision. Exclusion criteria EC4 and EC5 relate to our research goal. For instance, many articles were found to describe courses in computer engineering or mini-projects conducted prior to SE capstones which meant that they were out of the scope of this research and excluded based on EC4. Most of the papers left out during this stage met EC4. As for EC5, some studies were, for example, found to describe a whole curriculum with capstone courses playing only a minor part in the research, and they could therefore not provide answers to our research questions. A number of studies also evaluated a tool, method or framework relevant to the software engineering industry, not the capstone course itself. In many of these studies, the capstone course presented the researchers merely a convenient way of gaining study participants, which is why they did not fit this research. This led them to be excluded due to EC5.

\subsubsection{The third stage - Removal of duplicates and studies of poor quality}\label{third_stage}

The third stage was included mainly to rule out any duplicate data and studies of poor quality from our research. After the third stage, the final set of 127 papers remained.

\textbf{Duplicate data} -- All 171 primary studies remaining after the second stage describe real-life software engineering capstones. As educators often like to modify their courses over time to find the best ways of teaching, studies here too reflect on the changes done to the courses. Some authors also have written multiple articles based on the same capstone course. In such cases, the most recent article was chosen. Similarly, if a study describes several instances of the course in one paper, the principal characteristics of the most recent course instance were chosen for the data extraction. Choosing the latest instance of each course stems from the goal of this research to synthesise the current state of knowledge on capstone implementations. In addition, the decision of whether two descriptions of the same course are different enough for them to be included as their own capstone courses would have been too ambiguous and open for interpretation. \cite{kitchenham2007guidelines} also state that it is important not to include multiple publications of the same data, as it would seriously bias any results. Due to this procedure, 42 studies were removed from the final set.

\textbf{Quality assessment}\label{quality_assessment} -- In addition to inclusion/exclusion criteria, \cite{kitchenham2007guidelines} state that it is critical to perform a quality assessment on the primary studies. We also conducted a such assessment and used it to ensure that our final data set is of sufficient quality. At this stage, two studies were filtered out. Any tables or graphs presented from here on do not include these two excluded studies, and therefore represent the final set of 127 studies.

Table \ref{tab:quality_assessment} lists the set of questions used by \cite{dybaa2008empirical, ali2010systematic, mahdavi2013variability} which we also used to determine the quality of primary studies. Originally the questions were supposed to be graded on a dichotomous (``Yes`` = 1 or ``No`` = 0) scale \citep{dybaa2008empirical}, but we decided to use a three-point scale of ``Yes`` (= 1), ``To some extent`` (= 0.5) and ``No`` (= 0). This three-point scale has also been adopted by \cite{ali2010systematic} and \cite{mahdavi2013variability} and allowed us better to assess the studies where authors only provided some answers to the question. The two articles filtered out had a quality score of less than 4.

\begin{figure*}[hb!]
    \centering
    \includegraphics[width=1\linewidth]{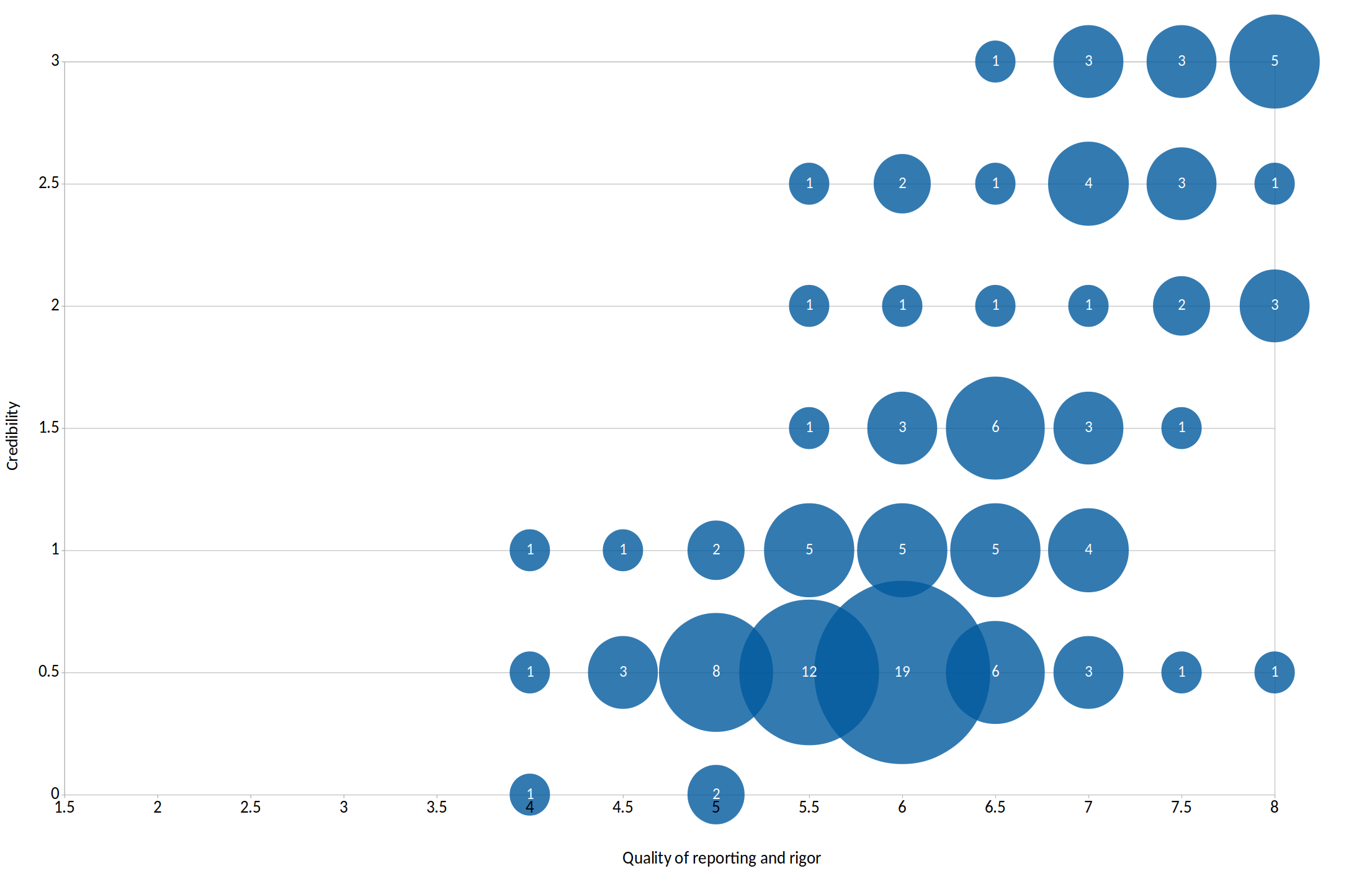}
    \caption{Quality scores of the final set of studies}
    \label{fig:quality_scores}
\end{figure*}

\begin{table}[width=.9\linewidth,cols=2]
\scriptsize
\caption{Questions for quality assessment}\label{tab:quality_assessment}
\begin{tabular}{p{0.05\linewidth}p{0.8\linewidth}}
\toprule
No. & Question \\
\midrule
Q1 & Is there a rationale for why the study was undertaken? \\
Q2 & Is there an adequate description of the context
(e.g.\ industry, laboratory setting, products used, etc.) in which the research was
carried out? \\
Q3 & Is there a justification and description for the research design? \\
Q4 & Has the researcher explained how the study sample (participants or
cases) was identified and selected, and what was the justification for
such selection? \\
Q5 & Is it clear how the data was collected (e.g.\ through interviews, forms,
observation, tools, etc.)? \\
Q6 & Does the study provide a description and justification of the data
analysis approaches? \\
Q7 & Has ‘sufficient’ data been presented to support the findings? \\
Q8 & Is there a clear statement of the findings? \\
\midrule
Q9 & Did the researcher critically examine their own role, potential bias and
influence during the formulation of research questions, sample
recruitment, data collection, and analysis and selection of data for
presentation? \\
Q10 & Do the authors discuss the credibility of their findings? \\
Q11 & Are limitations of the study discussed explicitly? \\

\bottomrule
\end{tabular}
\end{table}

In our assessment, we decided to group the first eight questions to represent the quality of reporting and rigour of the studies and final three questions to represent the credibility of evidence, similarly to \cite{ali2010systematic}. The grouped scores are presented in Fig. \ref{fig:quality_scores} and individual scores for each study can be found at \href{https://github.com/UniversityOfHelsinkiCS/article-additions/blob/main/capstone-courses-in-software-engineering/Quality_Assessment_Scores.png}{https://github.com/article-additions}. Regarding the quality of reporting, the selected primary studies performed fairly well. It was mostly clear how the data had been collected, and the relevance of the study was explicitly discussed. However, the aspect most studies were lacking was providing justifications either for the sample selection or research designs. Regarding the credibility of evidence, the studies performed fairly poorly. Interestingly, many of the otherwise well-established studies did not include a section for explicitly discussing the limitations of the study or the author's role in data and sample selection. This is indicated in the low averages of the credibility category. 

\subsubsection{Overview of the final papers}\label{overview_final_papers}

\begin{figure}[h]
    \centering
    \includegraphics[width=1\linewidth]{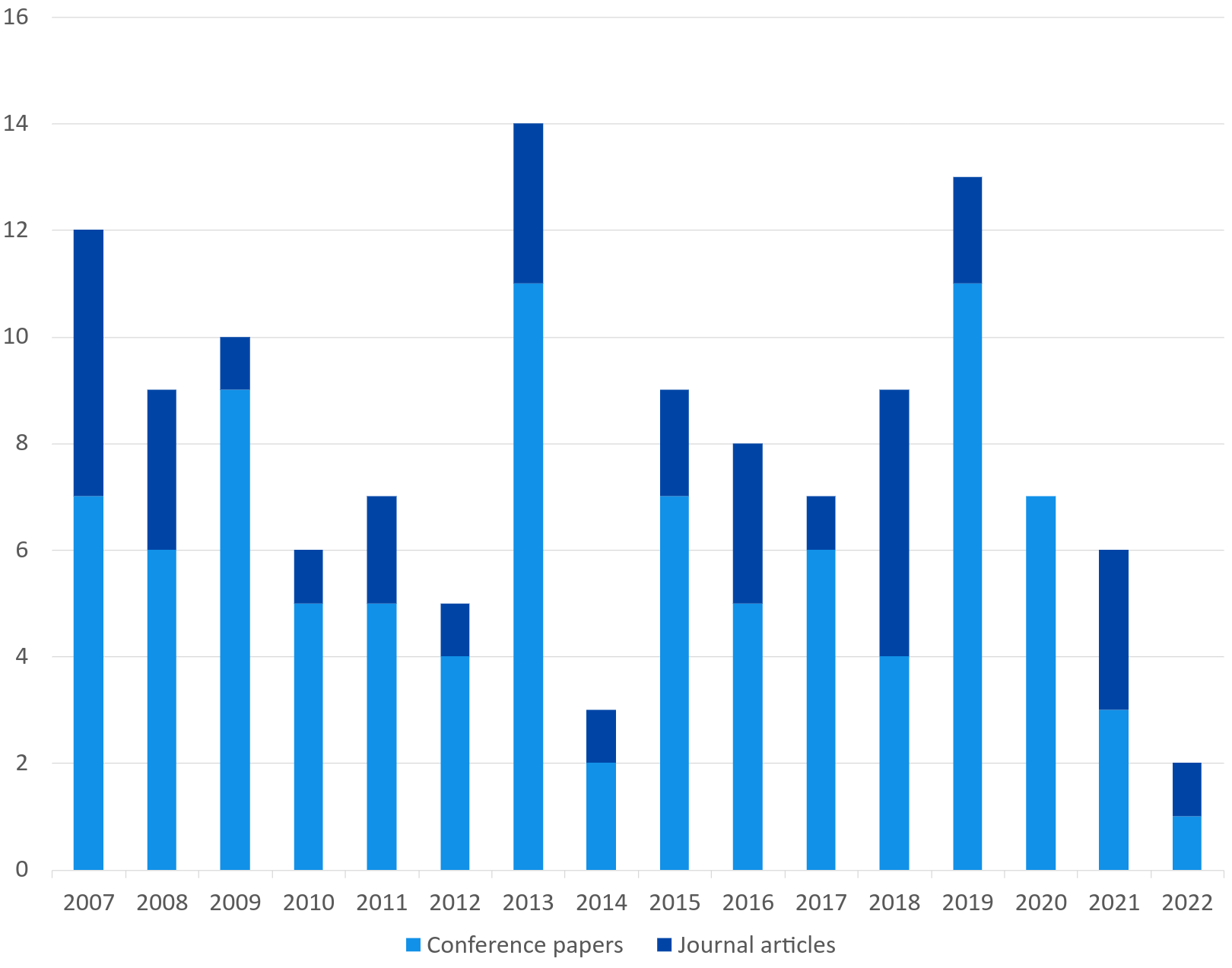}
    \caption{Timeline and types of primary studies}
    \label{fig:years_of_studies}
\end{figure}

The three stages taken resulted in 127 primary studies, published between 2007 and June 2022. Research activity in this area has been fairly steady over the years, as depicted in Figure \ref{fig:years_of_studies}. It is worth noting, that we searched for papers in June 2022, making the study amount for 2022 partial. Also, as explained in Section \ref{third_stage}, 42 earlier studies, which otherwise would have been valid for this research, were excluded from the final set of papers as there was a newer study of the same course available. This procedure skews the year distribution towards the end of the scale. The figure also shows the distribution by study type. In total, of studies 73\% were conference proceedings, and 27\% of studies were published as journal articles.

All the articles included for further analysis are listed in Appendix A, Table \ref{tab:final_sources}, and referenced later in this section with their publication ID in the table (i.e., S1--S127.) 

\subsection{Data extraction and synthesis}\label{data_extraction}

After applying the study selection process, the properties presented in Table \ref{tab:data_extraction} were extracted from the remaining 127 studies to a common datasheet. Table \ref{tab:data_extraction} defines how each extracted field relates to the research questions of this study. 

\begin{table}[width=.9\linewidth,cols=3,pos=ht!]
\caption{Data extraction form}\label{tab:data_extraction}
\begin{tabular*}{\tblwidth}{@{} LLL@{} }
\toprule
Identifier & Field & RQ \\
\midrule
F1 & Title & metadata \\
F2 & Author(s) & metadata  \\
F3 & Year & metadata \\
F4 & Publication venue & metadata  \\
F5 & Duration of the course & RQ1  \\
F6 & Course workload & RQ1 \\
F7 & Team sizes & RQ2 \\
F8 & Clients & RQ3.1  \\
F9 & Project sources & RQ3.2 \\
F10 & Artefacts produced & RQ4.1 \\
F11 & Project phases & RQ4.2 \\
F12 & Technologies & RQ4.3 \\
F13 & Student assessment & RQ5 \\
F14 & Outcomes of the course & RQ1--RQ5  \\
F15 & Quality score & QA \\
\bottomrule
\end{tabular*}
\end{table}

Values F1--F4 were extracted for basic documentation purposes. Items F5--F14 concern the course and its organisation presented in the study. Two of the studies present multiple separate capstone courses from different institutions. For these two studies (S8 and S93 in Table \ref{tab:final_sources}), the items F5--F14 were extracted for each of the courses they presented. For F5--F14, we were not only interested in quantifying these characteristics into statistics but also in providing implications of different course design choices. Therefore, if the study stated, for example, that they had a two-semester capstone course because it provided students adequate time to learn, we recorded both of these information pieces: the quantifiable duration as well as any such insight relating to the characteristic. This enabled us to analyse and discuss the course characteristics better in Section \ref{results}. A data-driven thematic analysis was applied to synthesise the qualitative data extracted as part of F5--F14 \citep{castleberry2018thematic}.

F5 and F6 were considered essential in assessing the general workload and duration of the course from the student's perspective. Few sources have given the duration of their course (F5) as months or weeks. These were rounded to the nearest amount of semesters. Courses lasting less than 4 months were categorised as ``less than one semester``, 4 to 6 months as ``one semester`` and anything more than 6 but less than 10 months as ``two semesters``. 

Team sizes (F7) included the number of students per team. The courses were also examined on whether the projects in the course were done for a client (F8). The client could be external to the course staff, the role of a client could be played by the course staff, and some projects did not have clients at all. Some sources present a mix of these categories, in which case, the source was labelled by the client category, which we thought was the most prevalent in the course.

We were also interested in how the project topics were generated (F9). Three main sources for projects were identified during the data extraction: course staff, external clients and the students themselves. The projects were also found to vary regarding whether the students were working on the same project idea or whether each team had their own initial problem to solve.

We extracted all the artefacts that students were expected to produce during the course (F10). This included both the deliverables used for grading the course as well as artefacts produced for project management reasons, as in most cases, it was hard to draw a distinction between the two. Evidence of project phases (F11) was extracted to find out which software life-cycle activities are gone through in these courses. F12 describes the technologies used in the course. We found that most of the studies do not explicitly specify all the technologies used for the projects in their courses, and moreover, these technologies could potentially include any software technologies available. We, therefore, categorised these into two categories, based on whether the main technology selections are made team-wise, or all use a common technology stack.

We extracted information on how the students learning process was assessed and improved throughout the course and how the student's progress and achievement were assessed at the end of the course (F13). The key outcomes in the study (F14) were also extracted to assess the advantages and disadvantages of the presented capstone characteristics. F15 was extracted as presented in Section \ref{quality_assessment} for quality assessment and study filtering.

\section{Results and analysis}\label{results}

This section represents quantitative statistics and qualitative outcomes of capstone characteristics extracted from the primary studies. The characterisation enables us to answer our research questions and ultimately helps educators when they are planning their capstone courses. 

\subsection{Duration (RQ1)}\label{duration}

\begin{table*}[width=1\textwidth,cols=4,pos=hb!]
\renewcommand{\arraystretch}{1.3}
\caption{Duration of capstone courses}\label{tab:duration}
\begin{tabular}{p{0.22\textwidth}p{0.15\textwidth}p{0.08\textwidth}p{0.45\textwidth}}
\toprule
Category & Number of studies & Percentage & Study identifiers \\
\midrule
Less than one semester & 10 & 8\% &
S16, S18, S40, S46, S55, S61, S78, S94, S99, S113 \\
One semester & 87 & 66\% & 
S1, S2, S3, S4, S5, S8b, S8d, S9, S11, S13, 
S15, S20, S21, S22, S23, S24, S25, S26, S27, S30,
S31, S32, S34, S35, S36, S37, S38, S42, S43, S44,
S45, S47, S48, S51, S53, S54, S56, S57, S58, S60,
S62, S63, S64, S65, S66, S67, S68, S69, S70, S73,
S74, S75, S76, S77, S79, S80, S81, S82, S83, S84,
S86, S89, S90, S92, S93a, S93b, S95, S97, S100, S102,
S103, S104, S105, S106, S107, S112, S115, S116, S117, S119,
S120, S121, S123, S124, S125, S126, S127 \\
Two semesters & 32 & 24\% & 
S6, S7, S8a, S8c, S10, S12, S14, S17, S19, S28,
S33, S39, S41, S50, S52, S59, S71, S72, S85, S87,
S88, S91, S96, S98, S101, S108, S109, S110, S111, S114,
S118, S122 \\
More than two semesters & 1 & 1\% & S49 \\
Not specified & 1 & 1\% & S29 \\
\bottomrule
\end{tabular}
\end{table*}

Regarding the actual course characteristics, we first looked into the reported duration of these courses (F5). A clear majority of institutions conduct capstone courses that last one semester (Table \ref{tab:duration}). Interestingly, this is in conflict with the \cite{ieeeSE} recommendations for undergraduate capstone courses, which propose having capstones lasting the entire academic year. However, the unfortunate reality is that not all curricula can absorb a full-year implementation [S117]. The capstone courses often are very labour-intensive for the teaching staff, with many teams to manage and evaluate throughout the projects [S39], [S112]. Students might have full- or part-time work, which makes the longer courses harder to arrange [S49], [S58], [S73]. Students also perceive two-semester capstone courses as laborious [S73], [S109], and some even the one-semester ones [S23], [S41]. In order to provide an intensive and realistic experience, many of these courses take up at least half a work-week [S18], [S28], [S41], [S69], [S73], [S109], [S127]. This again might make other courses taken simultaneously suffer [S41], which limits the possibilities for an intensive, year-long capstone.

However, the educators who had experiences with both, shorter and longer duration, had shifted to the longer duration since they felt it was impossible to reach the wanted depth in just a few months. S6 describes how they switched to a two-semester capstone as they found the one-semester projects inadequate in skill coverage and depth. S70 is written by students of the course, and they strongly recommend that their course be lengthened into one academic year from the current duration of one semester. S33 have had experiences with one-period, i.e.\ a quarter of an academic year, and three-period courses, and stated that the change to a longer version received overwhelmingly positive feedback from all the participating parties. Students were able to gain more hands-on experience in applying new and familiar tools and project management. Additionally, they learned to act when faced with unanticipated events as the teams experienced surprises -- regarding both technologies and people -- multiple times during the year. Industrial clients received more ambitious and polished products as a result of the course, and the course staff felt that the learning objectives for the course were finally truly met.

\subsection{Team sizes (RQ2)}\label{team_sizes}

\begin{figure*}
    \centering
    \includegraphics[width=1\textwidth]{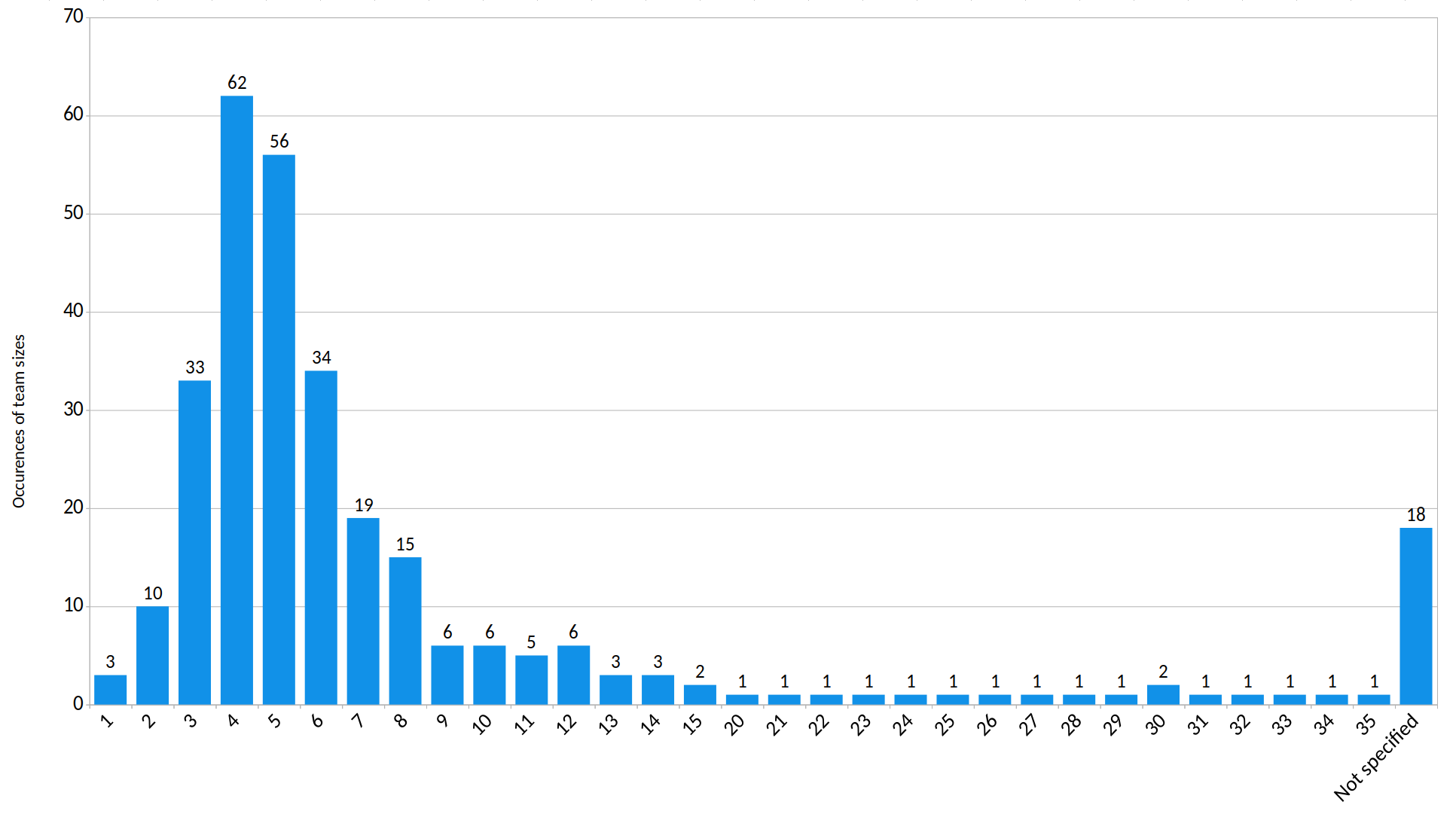}
    \caption{Team sizes in capstone courses}
    \label{fig:team_sizes}
\end{figure*}

To find out how many students there generally are in a project group, we extracted the reported team sizes (F7) for each course in Fig. \ref{fig:team_sizes}. If a study refers to their course having teams of 4--5 students, this is thus reflected in both columns 4 and 5. By looking at Fig. \ref{fig:team_sizes}, it is evident that capstone courses are almost always conducted as group projects. Only three institutions in our research allow their capstone or senior project courses to be completed as single-student endeavours [S11], [S89], [S111]. 

Team sizes vary a great deal, ranging from 1 to 35. Research has found that in very small groups, e.g., 2--3 students, the teams are unlikely to generate the dynamics and issues that are common in collaborative software development [S36], [S53], [S56], [S58]. Such a small team size does not present enough of a challenge [S36], and smaller groups are unable to complete substantial projects in a typical one-semester course [S53]. Having very small teams might also be unmaintainable in large programs with hundreds of, or even a thousand, students due to the extra organisational overhead each team causes [S19]. Going to the other extreme, larger groups with 7 or more students have often been found to be facing other kinds of problems, such as the inability to meet all together and other management and coordination issues [S30], [S36], [S39], [S53], [S56], [S78]. ``Free-rider`` problem is also reportedly common in larger teams, where it is possible for few students to take the bigger responsibility for ensuring the overall success, and the small contribution of others might go unnoticed [S9], [S56], [S58], [S69], [S87], [S121]. In larger teams, ensuring fair grading and an equal balance of work and responsibilities requires more attention from the course staff [S56], [S87]. 

The course conducted in S106 had one of the largest team sizes found in our research. The course had 15 students, all working on the same game project in one team. The idea was to simulate what large-scale game development in a diverse team feels like and what it takes to create production-quality games. The authors share that their approach was not entirely successful. In the aftermath of the course, it came up that some students wanted explicit direction while others felt that they wanted more autonomy and control. According to the authors, for the latter group of students, it was clear that they were uncomfortable following the leadership of the vision team and would have preferred to work on a project of their own design. However, the authors also mention that getting to work with your own project vision is a very unlikely case for any recent graduate, which is why they did try to come up with such a real-world teamwork scenario.

The majority of educators do seem to opt for the middle ground regarding team sizes and have 4 to 5 people working in a single group (Fig. \ref{fig:team_sizes}). This size is perceived as the sweet spot, cancelling out the negatives of the two extremes [S36], [S52], [S53], [S56], [S58]. Students themselves have also reported being satisfied with such a team size [S56]. Additional measures for combating any non-productive and opportunistic group behaviour, such as social loafing and free riding, have also been proposed. Conducting peer reviews has been proven to mitigate the risk of such behaviour [S72], [S98]. Some periodic monitoring should also be done by the course staff to ensure working team dynamics [S69]. Both of these will be discussed further in Section \ref{assessment_of_students}.

\subsection{Clients and project ideas (RQ3)}\label{clients_project_sources}

\subsubsection{Clients (RQ3.1)}\label{clients}

\begin{table*}[width=1\textwidth,cols=4,pos=ht!]
\renewcommand{\arraystretch}{1.3}
\caption{Clients of capstone courses}\label{tab:clients}
\begin{tabular}{p{0.22\textwidth}p{0.15\textwidth}p{0.08\textwidth}p{0.45\textwidth}}
\toprule
Category & Number of studies & Percentage & Study identifiers \\
\midrule
Clients external to the course staff & 76 & 58\% &
S2, S3, S4, S5, S7, S8a, S8d, S9, S10, S11,
S14, S15, S17, S18, S19, S23, S24, S27, S28, S29,
S30, S31, S33, S34, S35, S37, S38, S39, S41, S46,
S48, S49, S50, S52, S55, S58, S59, S60, S61, S62,
S63, S66, S67, S69, S71, S73, S78, S80, S81, S84,
S85, S86, S87, S88, S90, S91, S92, S93a, S93b, S94,
S96, S97, S98, S104, S108, S110, S112, S113, S114, S117,
S120, S121, S122, S124, S126, S127 \\
Course staff acts as clients & 16 & 12\% &
S1, S6, S12, S21, S40, S43, S53, S54, S57, S68,
S76, S82, S99, S115, S116, S123 \\
No clients & 38 & 29\% &
S8b, S8c, S13, S20, S22, S25, S26, S32, S36, S42,
S44, S45, S47, S51, S56, S64, S65, S70, S72, S74,
S75, S77, S79, S83, S89, S95, S100, S101, S102, S103,
S105, S106, S107, S109, S111, S118, S119, S125 \\
Not specified & 1 & 1\% & S16 \\
\bottomrule
\end{tabular}
\end{table*}

We also looked at who is in the role of a client for these projects (F8) and how the project ideas are sourced (F9). Almost half of the studies (42 \%) report conducting their capstone courses without clients that are external to the course (Table \ref{tab:clients}). In these courses, the course staff may act as clients or Product Owners for the projects or alternatively, the student teams work on their own and only report progress regularly to the course staff. S6 explains that they have instructors playing clients due to the difficulty of finding suitable clients. Being a small program in a rural institution makes the businesses and organisations suited for such collaboration far and scarce. Also, the institution's wish to own the intellectual property rights for the developed products puts off potential clients. For institutions, that would have suitable clients available, there is always the upfront investment in time and effort that the course staff has to make to contact said clients, guiding them through creating project proposals and assigning the students to these projects [S118]. S42 aimed to create a course with students from five different technical and non-technical disciplines, such as computer science and business informatics. S42 mentions that they need to be careful in how they organise the course so that it would suit the needs of all disciplines. Bringing an external client into the mix might not fulfil the learning goals for all students. Some studies also explain how the course outcomes are less predictable with multiple external clients. S114 have experienced several cases when the project sponsors did not show up for the bi-weekly meetings with the students. Such client behaviour caused very low motivation in the student teams, and some capstone projects failed due to client unavailability. S3, S23, S78 and S122 have made similar observations and stress the importance of finding committed clients to ensure a good experience for the students.

Despite these risks having real, external clients other than the course supervisor for the projects is recommended for both undergraduate and graduate capstones \citep{ieeeSE, ieeeSEGrad}. These clients can be from other units within the university [S7], [S9], [S14], [S52], [S58], [S86], [S87], [S94], local businesses [S7], [S9], [S86], [S87], [S98], [S110], [S122] or various non-profit organisations [S7], [S9], [S11], [S16], [S58], [S110]. Graduates of the program who already work in the industry are also a convenient way to find clients [S35]. Working closely with real-world clients has also often received highly positive feedback from students [S14], [S15], [S52], [S73], [S98], [S117] and organising staff alike [S14], [S35], [S84], [S98], [S117]. It has been found to increase the motivation and commitment of students, when there is an actual client with a real need behind the project [S9], [S14], [S15], [S35], [S66], [S73], [S84], [S121]. It has helped to keep the experience more realistic and credible in the students' eyes [S19]. Having industry clients improves the students' technical and nontechnical skills and better prepares them for the challenges they will face in the work-life [S14].

The collaboration has been reported to have benefits for the client too. S35 conducted a study to find out the reasons why clients participate in such project courses. The reasons included getting a tailored software product, researching new technologies and, as a clear number one, recruitment. Recruiting students could happen directly from the team or more indirectly by adding visibility among the students as potential employers. Others have noticed this benefit, too; it is not uncommon for students to get hired by the industry partner who sponsored their capstone project [S15], [S28], [S35], [S73], [S84], [S114], [S121]. S73 report having at least 60 out of a few hundred students gaining full- or part-time job offers based on the capstone project outcomes, in mere few years. Keeping the experience positive also for the clients, might make them come back with further project ideas [S35]. This helps to reduce the client acquisition overhead for years to come. Some organising institutions have even managed to attract more external clients than there are student teams, which has enabled them to collect a small fee from the ones participating in the course [S35], [S121].

\subsubsection{Project sources (RQ3.2)}\label{project_sources}

\begin{table*}[width=1\textwidth,cols=4,pos=ht!]
\renewcommand{\arraystretch}{1.3}
\caption{Sources for projects in capstone courses}\label{tab:project_sources}
\begin{tabular}{p{0.22\textwidth}p{0.15\textwidth}p{0.08\textwidth}p{0.45\textwidth}}
\toprule
Category & Number of courses & Percentage & Study identifiers \\
\midrule
External stakeholders propose project ideas & 81 & 62\% &
S2, S3, S4, S5, S7, S8a, S8b, S8c, S8d, S9, 
S10, S11, S14, S15, S17, S18, S19, S23, S24, S27, 
S28, S29, S30, S31, S33, S34, S35, S37, S38, S39,
S41, S46, S48, S49, S50, S52, S55, S58, S59, S60, 
S61, S62, S63, S66, S67, S69, S71, S72, S73, S77,
S78, S80, S81, S84, S85, S86, S87, S88, S90, S91,
S92, S93a, S93b, S94, S96, S97, S98, S104, S108, S110,
S112, S113, S114, S117, S120, S121, S122, S124, S125, S126,
S127 \\
Course staff provides project ideas & 27 & 21\% &
S1, S6, S12, S13, S16, S20, S21, S25, S40, S42,
S43, S45, S54, S57, S65, S68, S74, S75, S79, S82,
S99, S105, S106, S109, S115, S116, S123 \\
Students generate their own project ideas & 22 & 17\% &
S22, S26, S36, S44, S47, S51, S53, S56, S64, S70,
S76, S83, S89, S95, S100, S101, S102, S103, S107, S111,
S118, S119  \\
Not specified & 1 & 1\% & S32 \\
\bottomrule
\end{tabular}
\end{table*}

In addition to finding out the clients for these projects, we also looked into how the projects for these courses are sourced (F9). Three main ways for project sourcing were identified (Table \ref{tab:project_sources}). As the majority of courses have multiple external clients, the project ideas in these courses are mainly derived from the needs of the customer. In these cases, the organising staff often performs some pre-screening and scoping in collaboration with the clients, to ensure that the expectations for the projects are realistic and that the project scopes suit the intended learning outcomes [S9], [S24], [S35], [S52], [S61], [S87], [S90], [S94], [S98], [S121]. Capstone projects should generally not be on the critical path of any external organisation, as the course is intended to remain a safe learning place for the students [S35], [S52], [S87], [S90], [S121]. Some studies also emphasise that students are not supposed to be working for these clients, but be in collaboration with them [S9], [S52], [S87], [S121]. Thus, the projects need to remain such that the students can have a say in how the project will be developed. 

For 17\% of the courses, the students themselves are the main source for project ideas. These studies state that students are more motivated if they get to choose the project idea rather than have teachers assigning the projects [S36], [S53]. According to S53, if the team selects and defines the projects, their level of commitment and excitement to the project rises as the software system grows. At the end of the semester, the students have a strong sense of ownership towards the project, rather than feeling that they have just done one additional assignment [S36], [S53]. However, there are some potential pitfalls with this approach that educators should be aware of. S114 states that students should not be allowed to bring project ideas from the companies they work at or from their own businesses. S114 have found that it causes a conflict of interests for the student with the proposal and creates an unfair situation for the rest of the team. S36 lets students form their own teams and generate their own project ideas, but states this might not accurately reflect the situation in the students' future professional lives. If the project idea comes from the team itself, all complexities associated with requirements elicitation and analysis are eliminated [S73], making the experience less realistic. Real-life projects come with challenges relating to contradicting expectations coming from various external and internal stakeholders [S73].

For 21\% of the courses, the course staff provides the project specifications. Some educators have assigned the same project idea to all the student teams [S40], [S45], [S72], [S82] or even in some cases, all the students work on the exact same project in one team [S106], [S112]. Having the same project has the benefit of giving the course staff a consistent basis for grading and teaching [S40], [S59], [S72] and providing technical assistance to the students [S40], [S53]. In such cases, all teams will need to deal with the same complexities, project management issues and technology demands as in a typically constructed course, which makes the experience more predictable [S72]. Having one project idea also opens up the possibility for competition amongst the teams, e.g., which team will create the best design and implementation [S5], [S30], [S37], [S42], [S54], [S72], [S106] potentially even for an external client [S30]. It also possibly allows the course to focus more on the quality of the developed software [S5]. S5 has experience with both approaches, having multiple project ideas and having only one project. They have found it more productive and rewarding to focus on doing one project really well rather than juggling multiple projects and obtaining partial results.

\subsection{Project implementation (RQ4)}\label{projects}

We also aimed to uncover information on how capstone projects are implemented and what students are expected to do in these courses. For this, we extracted various information on the project implementation (F10--F12). 

\subsubsection{Produced artifacts (RQ4.1)}\label{produced_artifacts}

\begin{figure*}
    \centering
    \includegraphics[width=1\textwidth]{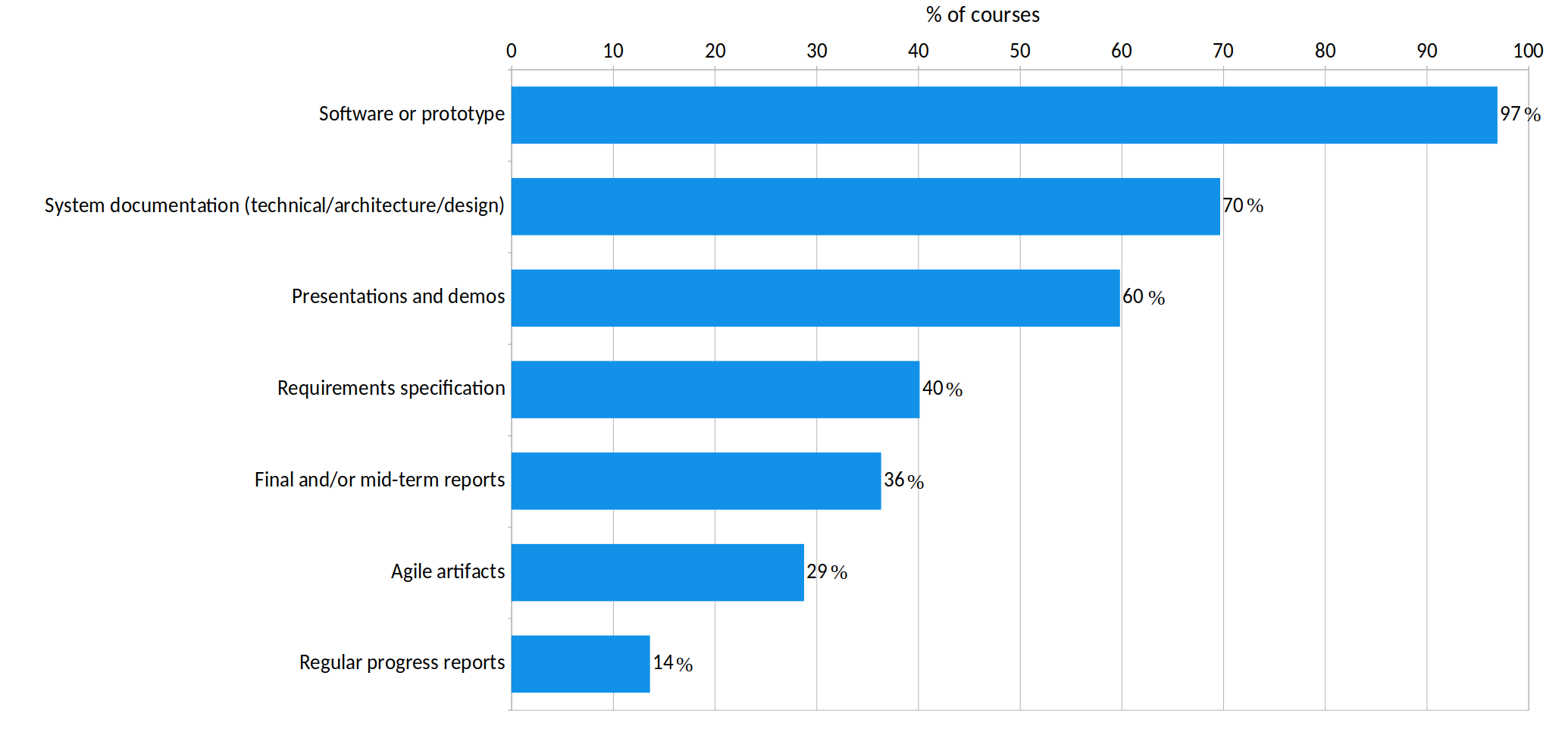}
    \caption{Most commonly mentioned artefacts produced by students in capstone courses}
    \label{fig:produced_artifacts}
\end{figure*}

We were interested in finding out what kind of artefacts students are expected to produce on the course (F10). It was often difficult to determine which artefacts were used for the final student assessment (i.e.\ grading) and which were only produced to manage the project in some way. Therefore we could not produce a list specifically of graded artefacts. Table \ref{fig:produced_artifacts} provides a rough view of the explicitly mentioned artefacts that students are expected to produce.

We were, however, able to determine that all but three courses [S16], [S17], [S103] expect some form of a software prototype, a software product or source code as the end deliverable. Out of the courses which did not include producing software, S16 conducted a course where the process and end deliverables are focused on students completing various research items, such as testing new team-based technologies (e.g.\ pair programming). For S17 the end deliverable was often a solution proposal for the client organisation's IT department, and S103 delivered an inter-disciplinary course which might result in software deliverables or alternatively in reports describing a software-based solution to the defined problems, e.g.\ using 3D engines for art.

Most studies mention requiring some documentation for the software project (Fig. \ref{fig:produced_artifacts}). The actual number of courses that require documentation might be larger than reported here, as we only counted the times the study explicitly mentions that project documentation is done on the course. Studies at the beginning of our time range are often following more ``plan-first`` software development approaches, such as the waterfall model, and as a result, the quantity and detail of non-software artefacts are substantial [S2], [S11], [S22], [S70]. The documentation usually starts heavily upfront by students producing project plans [S2], [S68], [S70], detailed designs [S2], [S22], [S52], [S68], architecture plans [S22], [S68] and test plans [S2], [S70]. In later, more agile, courses, there is less evidence of extensive documentation and planning. With agile projects, the system documentation is largely developed as the project evolves [S14], [S65], [S94]. Students in these courses often also produce agile artefacts, such as product backlogs, sprint backlogs and burndown charts for project management and planning purposes [S6], [S65], [S94], [S124]. 

A large share of the studies (60\%) explicitly mention that students must present or demonstrate their projects to wider audiences than just the immediate project team. This is a way to teach students how to present and explain their work also to a non-technical audience [S6]. The most common time for presentations is at the end of the semester, and typically these presentations are given at fairs or class sessions where all stakeholders of the course are invited [S2], [S4], [S6], [S33], [S95]. The format of final presentations varies from poster sessions [S6], [S28], [S34], [S44] to live demonstration sessions of the software [S89], [S95] to different kinds of demo videos [S1], [S73], [S118]. Students are sometimes also expected to draw up a project proposal or a pitch and then present it to the teachers or the class before starting to work on the implementation [S87]. In these cases, the purpose is to offer the students a chance to practice their pitching skills [S100].  

Software requirements are something that students are often expected to detail during the course. These can be written down as a Software Requirements Specification created before the implementation phase begins [S30]. For courses with agile methodologies, software requirements are often documented in an initial backlog with user stories, and the backlog is then updated as the project continues [S124]. Students are also quite often expected to write some form of a report at the end of the experience, either individually or as a group. The reports generally involve students reflecting on the learning done and development processes employed throughout the course [S6], [S19], [S21], [S24], [S27], [S41], [S63], [S65]. 

The balance between too little and too many other artefacts is a delicate one. Having more documentation and deliverables presents teachers with more opportunities to grade and assess students' understanding of software development processes [S110]. On the other hand, more documentation means less product which might not be in the interests of project clients [S49]. Indeed, some educators mandate only basic time tracking and reflective reporting from the students and have left the majority of the deliverables for the external client and team to decide [S24]. 

\subsubsection{Project phases (RQ4.2)}\label{software_lifecycle}

We sought evidence of the project phases and software life-cycle gone through on these courses (F11). Software life-cycle models include, with varying frequency and order, phases such as requirements gathering or solicitation, planning and designing, developing, testing and maintaining the product \citep{mishra2013comparative}. Out of the studies that discuss the development process and end product quality, the projects generally proceed from ideas to robust proof-of-concepts or products with few core requirements implemented [S1], [S6], [S7], [S11], [S12] [S16], [S18], [S20], [S21], [S22], [S26], [S40], [S42], [S45], [S52], [S54], [S55], [S62], [S64], [S65], [S68], [S79], [S70], [S72], [S75], [S77], [S78], [S87], [S90], [S94], [S95], [S99], [S100], [S106], [S107], [S109], [S115], [S118], [S122], [S123], [S124], [S125]. Students thus get to get experience the phases of planning, designing, developing and testing the products in these projects. In courses with clients, either external or internal, the students usually have to solicit the requirements from the clients (Table \ref{tab:project_sources}). However, sometimes the teachers provide students with ready-made feature or requirements lists [S12], [S21], [S45], [S79], [S82] and in some courses, students generate their own project proposals (Table \ref{project_sources}). The experience of requirements gathering is somewhat diminished in these cases.

Additionally, as the projects proceed from ideas to proofs-of-concept or simple, handed-off products, the projects generally do not include developing existing products, especially ones that are in production-use during the course. There are some courses where some of the projects have been production-ready at the end of the course, but these too were then handed over to the customer [S5], [S9], [S117]. This practice leaves students without the experience of working with existing products or products in the true maintenance phase of their software life-cycle. Assigning students to contribute to Free and Open Source Software (FOSS) projects is an emerging approach to remedy these shortcomings. The idea is to allow students to deal with existing codebases, often large and complex, such as the one they will face when working in the industry [S8b], [S8c], [S112], [S113]. Some courses have also had a continuation of earlier projects in the course to expose students to code generated by other people [S14], [S15], [S19], [S38], [103]. Both of these approaches allow students to maintain existing code, but they still present a minority in our research. 

\subsubsection{Project technologies (RQ4.3)}\label{technology_selection}

\begin{table*}[width=1\textwidth,cols=4,pos=h!]
\caption{Development technologies in capstone courses}\label{tab:technology_selection}
\begin{tabular}{p{0.22\textwidth}p{0.15\textwidth}p{0.08\textwidth}p{0.45\textwidth}}
\toprule
Category & Number of courses & Percentage & Study identifiers \\
\midrule
Projects use common technologies & 33 & 25\% & 
S1, S5, S8a, S8d, S11, S13, S22, S32, S37, S38,
S42, S43, S45, S46, S48, S53, S54, S57, S59, S72,
S76, S79, S82, S83, S93b, S99, S105, S106, S107, S109,
S112, S113, S124 \\
Choices done primarily team-wise & 65 & 50\% & 
S2, S3, S4, S8b, S8c, S9, S10, S12, S14, S19,
S24, S26, S29, S31, S33, S35, S36, S39, S44, S47,
S50, S51, S52, S56, S58, S61, S62, S65, S66, S68,
S69, S71, S73, S74, S75, S77, S80, S84, S85, S86,
S87, S88, S89, S90, S92, S95, S96, S98, S100, S101,
S102, S103, S104, S110, S111, S114, S116, S117, S118, S119,
S120, S121, S122, S125, S127 \\
Not specified & 33 & 25\% &
S6, S7, S15, S16, S17, S18, S20, S21, S23, S25,
S27, S28, S30, S34, S40, S41, S49, S55, S60, S63,
S64, S67, S70, S78, S81, S91, S93a, S94, S97, S108,
S115, S123, S126 \\ 
\bottomrule
\end{tabular}
\end{table*}

We also looked into the development technologies used in capstone courses and how they are selected (F12). Commonly in multi-customer courses or in courses with otherwise very differing project ideas, the technology choices are made based on the project (Table \ref{tab:technology_selection}). In these cases, the course staff does not impose an entirely common technology stack for all the projects. For some of these projects, the technology stack is based on the client's infrastructure [S35] and in some cases, the students get to make manager-like decisions on the suitable development technologies [S6], [S84]. Having the teams decide on the tools and technologies makes the students explore available options and justify their selections [S6], [S56], [S84]. S84 states that not only give them autonomy but also make them responsible for their own successes and failures. However, even though the majority of technologies would be selected based on the project and client, some studies recommend having some shared infrastructural tools and technologies [S12], [S19], [S36], [S65], [S102]. Version-control [S6], [S12], [S19], [S29], [S36], [S67], [S102], project management and communication tools [S19], [S29], [S65], [S67], [S80] and tools for continuous integration and delivery are examples of these [S32], [S78]. It has been found to make the management and evaluation of projects easier [S19], [S29]. Then again, having common development technologies for all projects is fairly common in cases, where the teachers provide students with the project requirements [S42]. In some cases, the evaluation methods focus heavily on the technical implementation, and the course graders might, for example, have sets of tests they like to run on each project to determine the quality [S109]. Some educators have the students compete on the same project proposal, which makes choosing a common stack justifiable [S37]. 

\subsection{Assessment of students (RQ5)}\label{assessment_of_students}

We were also interested in finding out how the assessment is conducted in these courses (F13), and to which extent students are given possibilities to reflect on their experiences. We looked at both the end-of-course student assessment (Section \ref{student_assessment}) and any continuous guidance and feedback students are given during the course (Section \ref{student_guidance}).

\subsubsection{End of course student assessment (RQ5.1)}\label{student_assessment}

All studies which explicitly described the course evaluation process had course teachers involved in it (Table \ref{tab:summative_assessors}). Teachers are generally the ones assessing any artefacts produced by students, such as the software product itself, reports and documentation done of the project [S118]. However, due to the shift from traditional development approaches to more agile ones, the quantity of delivered artefacts has decreased over time, leaving teachers with fewer data points for assessing students' comprehension of software development processes [S110]. In addition, many of the learning goals in capstone courses relate to soft skill development, such as the ability to work in a team or with a client and being able to manage a software project [S6], [S86], [S122]. These skills are generally employed when the teaching staff is not present, and teams work on the projects on their own, making the evaluation of these skills harder [S52], [S56], [S86]. For these reasons, several studies report having additional sources for student assessment beyond the teachers' evaluation of produced artefacts. Different kinds of (anonymous) peer evaluations are fairly common (31\%). They give course staff a look into the team dynamics during the course and help in detecting social loafing or free-rider behaviour [S12], [S86], [S112]. Similarly, self-evaluation is often done either in combination with peer evaluations [S56] or as a part of the reflection done in a project's final report [S65].

Some studies report utilising the client's opinion in the course assessment process. Clients can fill out a questionnaire considering each student's performance during the course [S86] or only evaluate the team's deliverables or presentations and their value from the client's point-of-view [S35], [S52], [S89], [S127]. As with self- and peer-reviews, educators use the client's opinion as a complementary source of assessment when grading students (Table \ref{tab:summative_assessors}). 

\begin{table*}[width=1\textwidth,cols=4,pos=ht!]
\renewcommand{\arraystretch}{1.3}
\caption{End of course assessment}\label{tab:summative_assessors}
\begin{tabular}{p{0.22\textwidth}p{0.15\textwidth}p{0.08\textwidth}p{0.45\textwidth}}
\toprule
Category & Number of courses & Percentage & Study identifiers \\
\midrule
Course staff & 93 & 71\% & 
S1, S2, S4, S6, S7, S8a, S8b, S8c, S8d, S9,
S12, S13, S17, S18, S19, S20, S22, S23, S24, S25,
S26, S27, S28, S29, S30, S31, S33, S35, S36, S37,
S39, S40, S41, S42, S43, S44, S46, S47, S48, S51,
S52, S53, S56, S58, S62, S63, S64, S66, S67, S68,
S69, S71, S72, S73, S74, S77, S80, S81, S82, S84,
S85, S86, S87, S88, S89, S90, S94, S95, S96, S97,
S98, S99, S100, S101, S102, S103, S104, S105, S106, S107,
S108, S109, S110, S111, S112, S116, S117, S118, S120, S123,
S124, S126, S127 \\
Students' peer-evaluations & 40 & 31\% &
S2, S4, S7, S8a, S8b, S12, S13, S27, S30, S31,
S33, S37, S41, S42, S46, S48, S51, S52, S56, S58,
S63, S64, S67, S72, S73, S74, S81, S84, S86, S87,
S90, S98, S106, S107, S108, S110, S112, S117, S124, S127 \\
Students' self-evaluations & 24 & 18\% &
S1, S2, S4, S22, S23, S27, S37, S41, S42, S47,
S51, S56, S62, S64, S69, S73, S74, S80, S81, S86,
S87, S116, S126, S127 \\
External project clients & 17 & 13\% & 
S4, S20, S24, S29, S33, S35, S37, S42, S58, S73,
S84, S85, S86, S89, S96, S98, S127 \\
Others & 1 & 1\% & S17 \\
Not specified & 38 & 29\% & 
S3, S5, S10, S11, S14, S15, S16, S21, S32, S34,
S38, S45, S49, S50, S54, S55, S57, S59, S60, S61,
S64, S65, S70, S75, S76, S78, S79, S83, S91, S92,
S93a, S93b, S113, S114, S115, S119, S121, S122 \\
\bottomrule
\end{tabular}
\end{table*}

\subsubsection{Continuous student assessment and guidance (RQ5.2)}\label{student_guidance}

\begin{table*}[width=1\textwidth,cols=4,pos=ht!]
\renewcommand{\arraystretch}{1.3}
\caption{Continuous student assessment and guidance}\label{tab:student_guidance}
\begin{tabular}{p{0.22\textwidth}p{0.15\textwidth}p{0.08\textwidth}p{0.45\textwidth}}
\toprule
Category & Number of courses & Percentage & Study identifiers \\
\midrule
Course staff & 99 & 76\% &
S2, S3, S4, S6, S8a, S8b, S8d, S9, S12, S15, S16,
S17, S18, S19, S20, S22, S23, S24, S25, S27, S28,
S29, S30, S31, S32, S35, S36, S37, S39, S40, S41,
S43, S44, S45, S46, S47, S48, S49, S50, S51, S52,
S53, S54, S55, S56, S57, S59, S60, S62, S63, S64,
S65, S66, S67, S68, S69, S71, S73, S74, S75, S76,
S77, S78, S80, S81, S82, S84, S85, S86, S87, S88,
S90, S91, S92, S93a, S94, S95, S96, S97, S98, S99,
S100, S101, S102, S103, S105, S106, S108, S109, S110, S112,
S113, S114, S115, S116, S122, S123, S124, S126, S127 \\
More experienced students & 23 & 18\% &
S5, S8a, S12, S14, S17, S19, S35, S40, S48, S58,
S61, S68, S81, S85, S92, S95, S104, S105, S112, S113,
S118, S119, S121  \\
Industry advisers (other than project clients) & 22 & 17\% &
S5, S8a, S8b, S8c, S10, S20, S25, S52, S58, S65,
S71, S72, S73, S77, S80, S83, S89, S93b, S98, S106,
S118, S120 \\
Not specified & 15 & 11\% &
S1, S7, S11, S13, S21, S26, S33, S34, S38, S42,
S70, S79, S107, S111, S117 \\
\bottomrule
\end{tabular}
\end{table*}

Many studies specifically mention that the teams should not be left entirely on their own to complete the course project and should be guided along the way [S6], [S9], [S10], [S16], [S18], [S53], [S69], [S72], [S86]. Three main ways for conducting such continuous assessment and guidance were found based on our research (Table \ref{tab:student_guidance}). Studies where there is no mention of the teacher, or anyone else, having an active role in how the teams work during the course, fall into the category ``Not specified``. If the course staff only passively receives reports of the student's progress and evaluates the course outcomes after its completion, these are not the active guidance of the teams we were looking for. Some courses have several types of guidance present, in which case the study has been listed under each corresponding category in Table \ref{tab:student_guidance}. Only 11\% of the studies do not explicitly specify having any ongoing feedback and guidance system present during the course.  

The most convenient way is to have the course staff, such as the responsible teacher or hired teaching assistants, acting in an advisory role (76\%). The intensity of the guidance given by course staff varies a great deal between these courses, or even within these courses. Sometimes course staff provides oversight in a more supervisory role and intervenes in the team's work if any conflicts arise or the team includes clearly non-contributing students [S6]. On the other end, some instructors have weekly meetings with the students where the teachers actively propose solutions and guide the teams with technical and non-technical issues and team dynamics [S6], [S88]. Some teachers prefer even to practically manage the team [S6]. S38 explains that the most successful changes made on the course were those that allowed the course staff to take a more active role in each team. The grades of students improved, and the teams were able to complete more functionality of the software products.

Another, often complementary, guidance form is to have industry experts occasionally participate in the course (16\%). This can be seen as especially relevant when the course projects are focused around a common theme, for instance, the gaming industry [S25], [S106]. However, finding the correct balance in this type of assessment, without a client relationship, has sometimes proven to be tricky. S106 had industry experts from the gaming and software industries participating as advisers on their course. The advisers' feedback on the students' game product was mainly positive and encouraging. While the staff took their remarks to mean that the game concept and development for the moment were commendable, the students took the feedback to mean that the prototype was, as presented, worthy of praise. This presented a dichotomy that never really resolved: students felt that the project was near-complete, whereas the instructors felt that the project was, at best, a rough sketch.

Many authors have noticed the upsides of having more experienced students outside of the course staff, mentoring the students in the course (18\%). These can be, for instance, students who have completed the project course themselves in the past year [S122]. This has been found to benefit both the project implementation and group dynamics: an active and knowledgeable coach can, for example, help students ask clarifying questions of the customer, overcoming fear of these being stupid and saving days or weeks [S122].

Interestingly, forming a capstone team of the final year students with similar skill levels are in accordance with the ACM/IEEE Curriculum Guidelines for Undergraduate SE Degree Programmes \citep{ieeeSE}, but leaves out an integral part of the real software development team experience: junior and senior positions. This discrepancy has been noted in some studies [S19], [S35], [S58], [S98], where the course implementation has gone beyond having senior students just as advisors. In these capstones, less-experienced students work as junior developers and more-experienced students as senior developers or team leaders. S19 organised their capstone course in a way that students are required to work two course units on the same project, one unit as a junior member and one unit as a senior member. Each unit lasts one period (a quarter of an academic year), but the periods do not have to be consecutive to allow some flexibility for students in organising their studies. In order for such an arrangement to work, the projects in the course are large, long-term products, which undergo enhancements over a number of semesters. S19 found that for junior students, this setup allowed a smooth transition to the project, up-skilling on relevant skills and acquiring the necessary orientation from senior students. Senior students, on the other hand, were enthusiastic about mentoring junior students and finding answers to their questions ranging from project requirements to the technology stack. S98 have similarly split their capstone project into two parts with junior and senior positions. They also had faculty mentors with industrial experience mentoring the student teams working with external clients. According to S98, having this course design enabled them to create an effective industrial simulation. S98 reports that students used tools and practices prevalent in the industry but frequently not taught in university and were able to develop professional and team working skills more intensively.

\section{Discussion}\label{discussion}

In this research, the main objective was to understand how tertiary education institutions conduct their SE capstone courses. This was done by looking at the characteristics of capstone courses through an extensive literature review. Firstly, we summarise the main findings of this study and compare them to the findings of previous systematic literature reviews, whenever appropriate (Section \ref{main_findings}). Secondly, we present suggestions for further research in this area in Section \ref{implications}. Finally, we discuss the validity of the results in Section \ref{validity}.

\subsection{Main findings}\label{main_findings}

\subsubsection{Duration (RQ1)}\label{findings_duration}

Despite \cite{ieeeSE} recommending that undergraduate SE capstones should span the whole academic year, most of courses identified here last only one semester. This is in line with findings regarding course models by \cite{dugan2011survey}. In our research, the studies presenting two-semester capstones often found one-semester courses inadequate in depth and breadth of skills they can provide. These studies reported that a longer course better prepares students for the experiences they can expect in their working life in software engineering. There are, however, some real-world constraints to why the courses generally are shorter, such as cramped curricula and the time and effort capstones require from both, the staff and the students. 

\subsubsection{Team sizes (RQ2)}\label{findings_team_sizes}

We found that capstone courses are generally conducted as large-scale group projects. The team sizes varied greatly between courses, ranging from 1 to 35 students in one team. \cite{dugan2011survey} found no agreement in the literature on the appropriate team sizes. Our results were somewhat contradictory to this. In our research, several studies that reported experiences with different team sizes had found the optimal to be 4--6 students per group. This was also reflected in average team sizes for capstones we found based on our research. For a group of 2--3 students, there is no communication challenge to solve, and smaller groups often are not able to accomplish larger projects. In contrast, larger teams often present too many issues for communication, coordination and fair grading of students. Larger teams also require more effort from the teaching staff to ensure an even distribution of work.  

\subsubsection{Clients and project ideas (RQ3)}\label{findings_clients}

In our research, 58\% of the studies reported having external clients for student projects (RQ3.1) and projects are based on the real needs of external stakeholders in 62\% of the courses (RQ3.2). Having clients outside the immediate course staff presents more work for the teachers, but is often rewarding for students when they get to work for real clients on real projects. The motivation boost in students, as well as the positive implications in their skills and employment after the course, were found to be among the top reasons for having real clients with real projects. Using external clients is also recommended for both undergraduate and graduate degree programmes in software engineering \citep{ieeeSE, ieeeSEGrad}. Despite these benefits, there still is a considerable number of capstone courses (42\%), where the course staff acts as the client for these projects, or there is no client for students to interact with regularly. In addition, there are quite many courses where the teacher provides the students with project specifications (21\%) or students themselves generate project proposals (17\%). Such courses generally are less burdensome for teachers who do not have to get involved in sourcing multiple clients and projects. Students are often also motivated when they get to choose a project topic of their own. However, in these cases, students do not get to experience requirement solicitation and planning the project with an external stakeholder, who might not be technically knowledgeable at all. The taxonomy used by \cite{dugan2011survey} relates to project topics and not particularly project sources or clients making it difficult to assess whether there has been changes in this over the years. Moreover, our work demonstrates that external stakeholders can get involved in many ways. In the future, it would be important to explore this in more detail, and evaluate the consequences as demonstrated by \citet{steghofer2018involving}. 

\subsubsection{Project implementation (RQ4)}\label{findings_project_implementation}

To understand the project implementation, we first looked into the artefacts that students are expected to produce throughout the course (RQ4.1). We found that in 97\% of the courses, students were involved in developing some form of a software product as an end deliverable. Additionally, students were often required to produce agile development artefacts (e.g.\ product and sprint backlogs), project plans and software documentation. The produced artefacts, therefore, supported the idea of a well-rounded software development experience. In our research, the role of documentation was slightly different from the role it had in the survey done by \cite{dugan2011survey}. In their classification, the core written documents involve project proposals, requirements documents, project plans, designs, test plans and user manuals. We, on the other hand, found that when courses had shifted towards more agile development approaches, the number of written assignments had reduced, and the documentation was being generated more throughout the course rather than as detailed plans up front.

Secondly, we investigated the phases that these projects generally go through (RQ4.2). We found that projects often proceeded from an idea or a ready-made list of requirements into project delivery. For a large number of courses, the students start from scratch and produce a prototype or a software product that is handed off to the clients or teachers at the end of the course. Therefore, the maintenance of existing software products and working with existing codebases is often not experienced. In contrast \cite{dugan2011survey} states that regardless of the software process model, the common phases were requirements, design, implementation, testing, presentation and maintenance. However, they do make the same finding we made that, maintenance was frequently mentioned in the literature with little supporting detail of its actual implementation. Maintenance thus still remains an issue that is left with little attention in SE capstone literature, despite its high relevance in the industry. Some educators have solved this by involving large projects, which go through various incremental improvements over the years in their courses. Others have students contributing to large Open Source projects, but both of these approaches were still found to present a minority. \cite{dugan2011survey} makes no remarks on Open Source projects being used as a solution to this problem like we did, which would indicate that it is an upcoming solution to this problem. Indeed, the studies in our research covering large Open Source projects were written after the survey by \cite{dugan2011survey}.

Finally, we investigated what technologies are used in these courses and how the selections are made (RQ4.3). We found that studies mostly do not explicitly describe all the technologies used in these courses. We also found that for the majority of courses, the technology selections are made based on the project specifications and needs. This often entails students learning new technologies and having to justify their selections.  

\subsubsection{Assessment of students (RQ5)}\label{findings_assessment}

Regarding end-of-course student assessment (RQ5.1) we aimed to find out what constitutes the course grade in the end. A large number of studies gave inconclusive answers in this regard and did not describe grading rubrics in detail. Therefore, we were unable to draw any conclusions about what artefacts or assignments formed the final grades. However, we were able to determine fairly well, who does the final assessment and how well students are given a chance to reflect on their experiences. In all studies that discussed the student assessment, the teachers were involved in determining the final grades. Any sort of concrete deliverables (produced software, plans, agile artefacts, reports) were generally graded by the teacher. These artefacts provided teachers with some understanding of how well students understood the phases of software development. A minority of studies also mention including students in the assessment process, in the form of self- and peer-reviews. These both have proven not only to hold the individual students more accountable during project work but also to give valuable insight for the teacher into the soft skill development of an individual student.

Continuous assessment and guidance during the course (RQ5.2) were explicitly addressed in most studies (89\%). The survey by \cite{trevisan2006review} focused on similar sort of assessment practices when they sought to find out, how often engineering capstones implement classroom assessment. They found only 32 articles in all engineering disciplines between the years 1994 and 2006 which discussed classroom assessment schemes. While their scope is tighter, it seems that the guidance of students during the course would have increased in the past 15 years or so. \cite{trevisan2006review} also reflects on this, stating that the importance of classroom assessment has gained traction in recent years. 

Our research showed that any sort of mentoring or coaching was found to be highly beneficial for the students. It increased the success rate of projects and helped teachers to identify problems early on. Course staff were the ones that most often guided students during their projects, and several courses had hired teaching assistants for such positions. Some studies also found that having more experienced students advising the capstone participants was a rewarding experience for both groups. Mentoring activities are also common in real-life companies where graduates generally join an existing team with various skill levels.

\subsection{Implications for practitioners and researchers}\label{implications}

A large amount of research found on software engineering capstones shows that capstones are a common way for educators to prepare students for varying aspects of working life. For an educator to find ways to implement a capstone course, it would be too a time-consuming task to go through all the published primary studies and distil the experience and evidence into concrete suggestions. For such situations, we have provided an overview of the most common characteristics of capstone courses, and what kinds of choices regarding each of these can be made.

The overriding guideline set by the \cite{ieeeSE} for undergraduate SE capstone courses is that they should help to ensure that the curriculum has a significant real-world basis. Capstones are expected to be the culminating experience that ties everything learned so far together and prepares students for the working life in software engineering \citep{ieeeSE, ieeeSEGrad}. Interestingly, our research revealed that primary studies on capstone courses only rarely included experiences, of course, alumni or industry clients. This would be important to see how well these courses reflect what students are expected to do in real life. We would like to see more research done on how well these courses capture what students face later on in their careers. It would also be worthwhile to see more controlled, comparative studies where one of the presented characteristics is changed and its impact on the course outcomes. Most of the research identified here does not provide controlled, comparative results on the capstone characteristics.

\subsection{Threats to validity}\label{validity}

\subsubsection{Deviations from the procedures for systematic reviews}\label{deviations}

Although we aimed to use the guidelines provided in \cite{kitchenham2007guidelines} to perform our systematic review, we had deviations from their procedures. In our research, the study selection and data extraction were carried out by the first author rather than by a group of researchers. This means that some relevant papers might have been excluded or that some of the collected data may be erroneous.

\subsubsection{Inaccuracy and bias in selected papers for review}\label{bias_selected}

One of the main limitations of any review is the possible bias in the study selection process. In our case, we included only studies considering software-related capstone courses with a relatively tight scope; for instance, we did not include any studies with courses on embedded systems or computer engineering. However, we were clear in our goals of describing only software capstones. A similar kind of search has also been conducted in the earlier survey done by \cite{dugan2011survey}, whose search strings were ``capstone`` and ``software engineering course`` into a selected set of journals in SEE and CS education. And to ensure that the selection process was as unbiased as possible, we described the employed search strategy and the inclusion and exclusion criteria in detail. This way, we aimed to make the selection process as visible to the reader as possible. 

The primary studies only represent the capstone courses with some aspects or outcomes worthy of publication. Therefore the study sample in our research might be skewed toward successful, well-planned courses or easy to research courses. We believe this is not a problem in describing how versatile the projects can be. However, quantitative aspects of the data (e.g., the portion of the courses having an external client) should be addressed with caution as certain kinds of courses may be more likely in real life than in SE education research. Finally, we also acknowledge that there are similar courses organised under other related disciplines, such as data science and computer engineering. We, however, knowingly chose to leave other disciplines out of the scope of this research as we wanted to provide a classification and insights specifically on software-related capstones. The \cite{ieeeSE} recommendations we derived our research questions on, were also provided specifically for software engineering capstones.

\subsubsection{Inaccuracy and bias in data extraction}\label{bias_data}

As with any systematic review, one of the main limitations is the potential bias and inaccuracy of the data extraction procedure. This is also the most likely step with inaccuracy in our research. For example, the quality assessment was done by the first author, whose interpretation of quality might differ from that of another researcher. The distinction between whether a study has explicitly discussed limitations (``Yes``) or they have only shortly referred to a limitation of the study (``Somewhat``) is something that another researcher might view differently. However, the two summed-up categories presented, rigour and credibility, aimed to diminish the impact of a single quality assessment question and evaluate the study rather as a whole. 

It is also worth noting that the primary studies presented in this review are not exclusively written to provide course descriptions or general course evaluations. Some studies have a section dedicated to the course overview, which might have provided all the details of the course structure we needed. Then again, some studies had the relevant details scattered across various sections and might not have been explicitly referred to as our categories suggest. Especially regarding the produced artefacts and student assessment, the descriptions varied greatly in terms of detail and clarity. In situations like these, some interpretation was needed. To mitigate this problem, we tried to keep the categories generic and descriptive, so that it would be easy to grasp the general outline of each course. We also refrained from reading too much into the text itself. For instance, if the study mentioned that the student teams were composed of ``at most 5 students``, we left these courses in the category ``not specified``. 

\subsubsection{Lack of third party assessment of capstone courses}\label{lack_of_third_party}

Usually, at least one of the authors of the study was somehow involved in organising the course in question. Additionally, quite a large portion of these reports lacked an honest evaluation of the author bias, as can be seen in Section \ref{quality_assessment}. Therefore there is an inherent lack of truly objective third-party assessment of these SE capstone courses in the literature. This is something that we were unable to affect but is worth noting. We would welcome more research on capstone courses, or on SE education in general, where the author is an unbiased third party.

\subsubsection{Evaluation of review}\label{review_evaluation}

For evaluating any SLR, \cite{kitchenham2009systematic} present criteria based on four quality assessment (QA) questions. We will briefly provide answers to each one of these.
 
QA1 --- Are the review's inclusion and exclusion criteria described and appropriate? Yes, we have explicitly defined and described the inclusion and exclusion criteria. The foundation for the criteria stems from our research objectives and aims to ensure that the studies included in the review are of sufficient quality and help to answer our research questions. 
 
QA2 -- Is the literature search likely to have covered all relevant studies? \cite{kitchenham2009systematic} state that if the authors have searched 4 or more digital libraries and included additional search strategies, this criterion is met. In this research, we did search 4 digital libraries and included a description of our search strategies, so this criterion is fulfilled. 
 
QA3 -- Did the reviewers assess the quality/validity of the included studies? We did use a question set used by many similar SLRs to assess the quality and validity of the included studies. Therefore this criterion is also met. 

QA4 -- Were basic data/studies adequately described? We provided bibliographical references to each of the studies used, described from various viewpoints the target of their research (i.e.\ the capstone course presented in each of them), described how the data was collected in each of the studies and synthesised the reported outcomes. Therefore it is safe to say, that this criterion was met as well. 

\section{Conclusions}\label{conclusions}

This research aimed to understand how software engineering capstone courses are organised in tertiary education institutions. For this purpose, we conducted a systematic literature review, including 127 primary studies on SE capstone courses. The characteristics were synthesised into a taxonomy consisting of duration, team sizes, clients and project sources, project implementation and student assessment. Based on the synthesised justifications and outcomes for these characteristics, we provided suggestions on how the courses can be organised and what the trade-offs are to be weighted regarding each characteristic. 

The main curriculum guideline that capstones should help to accomplish is ``The curriculum should have a significant real-world basis`` \citep{ieeeSE}. In our research, we focused on the concrete recommendations given to accomplish this goal and formulated our research questions based on them. We found out that the courses have a software implementation as the main deliverable, the students are assessed based on various factors, not just the delivery of a working system, and the projects in these courses are almost always completed as group assignments. Students were also often given guidance and continuous assessment throughout the course via written and oral feedback on their progress and deliverables. The area which educators should pay attention to is the duration of the course which in practice is one semester, whilst for instance, \cite{ieeeSE} recommends having two-semester courses to reach adequate depth and breadth in skills and experiences. A considerable number of courses also did not have a client external to the course staff, despite external clients being recommended for undergraduate and graduate capstones \citep{ieeeSE, ieeeSEGrad}. In these cases, the project specifications were generated by the course staff or the students themselves. Such arrangements tend to leave students without the experience of having to solicit, negotiate and implement requirements set by a real client. In addition, the projects usually progress from idea to product, and often do not include maintenance, especially that of pre-existing projects. These characteristics somewhat diminish the real-world compatibility of the course.


\bibliographystyle{cas-model2-names}
\bibliography{cas-refs}

\clearpage
\appendix

\section{Primary studies}

\begin{table}[width=1\textwidth,cols=5, pos=!b]
\begin{minipage}{\textwidth}
\centering
\scriptsize
\caption{Included sources for data extraction}\label{tab:final_sources}
\begin{tabular}{p{0.07\linewidth}p{0.19\linewidth}p{0.03\linewidth}p{0.31\linewidth}p{0.28\linewidth}}
\toprule
ID	& Author(s) & Year & Title & Source title \\
\midrule
S1 & Marzolo, P., Guazzaloca, M., Ciancarini, P. & 2021 & “Extreme Development” as a Means for Learning Agile & International Conference on Frontiers in Software Engineering	\\
S2 & Tan, J., Jones, M. & 2008 & A case study of classroom experience with client-based team projects & Journal of Computing Sciences in Colleges \\
S3 & Wong, W., Pepe, J., Stahl, J., Englander, I. & 2013 & A collaborative capstone to develop a mobile hospital clinic application through a student team competition & Information Systems Education Journal \\
S4 & Tappert, C. C., Stix, A. & 2011 & A decade review of a masters-level real-world-projects capstone course & Info. Systems Educators Conf., ISECON 2011 \\
S5 & Gotel, O., Kulkarni, V., Say, M., Scharff, C., Sunetnanta, T. & 2009 & A global and competition-Based model for fostering technical and soft skills in software engineering education & 22nd Conference on Software Engineering Education and Training, CSEE\&T 2009 \\
S6 & Scott, A., Kreahling, W., Holliday, M., Barlowe, S. & 2017 & A holistic capstone experience: Beyond technical ability & 18th Annual Conference on Information Technology Education	\\
S7 & Koolmanojwong, S., Boehm, B. & 2013 & A look at software engineering risks in a team project course & 26th International Conference on Software Engineering Education and Training, CSEE\&T 2013 \\
S8abcd & Braught, G., et al. & 2018 & A multi-institutional perspective on H/FOSS projects in the computing curriculum & ACM Transactions on Computing Education \\
S9 & Mertz, J., Quesenberry, J. & 2019 & A scalable model of community-based experiential learning through courses and international projects & 2018 World Engineering Education Forum - Global Engineering Deans Council, WEEF-GEDC 2018	\\
S10 & Bloomfield, A., Sherriff, M., Williams, K. & 2014 & A Service Learning Practicum capstone & 45th ACM technical symposium on Computer science education \\
S11 & Brazier, P., Garcia, A., Vaca, A. & 2007 & A software engineering senior design project inherited from a partially implemented software engineering class project & 37th Annual Frontiers in Education Conference - Global Engineering	\\
S12 & Morales-Trujillo, M.E., Galster, M., Gilson, F., Mathews, M. & 2021 & A Three-Year Study on Peer Evaluation in a Software Engineering Project Course & IEEE Transactions on Education	\\
S13 & Liang, Z., Chapa-Martell, M.A. & 2019 & A Top-Down Approach to Teaching Web Development in the Cloud & IEEE International Conference on Teaching, Assessment, and Learning for Engineering, TALE 2018	\\
S14 & Murphy, C., Sheth, S., Morton, S. & 2017 & A Two-Course Sequence of Real Projects for Real Customers & Conference on Integrating Technology into Computer Science Education, ITiCSE 2017	\\
S15 & Rusu, A., Rusu, A., Docimo, R., Santiago, C., Paglione, M. & 2009 & Academia-academia-industry collaborations on software engineering projects using local-remote teams & 40th ACM Technical Symposium on Computer Science Education, SIGCSE'09 \\
S16 & Stettina, C.J., Zhao, Z., Back, T., Katzy, B. & 2013 & Academic education of software engineering practices: towards planning and improving capstone courses based upon intensive coaching and team routines & 26th International Conference on Software Engineering Education and Training, CSEE\&T 2013 \\
S17 & Venson, E., Figueiredo, R., Silva, W., Ribeiro, L.C.M. & 2016 & Academy-industry collaboration and the effects of the involvement of undergraduate students in real world activities & IEEE Frontiers in Education Conference, FIE 2016	\\
S18 & Eloe, N., Hoot, C. & 2020 & Accommodating Shortened Term Lengths in a Capstone Course using Minimally Viable Prototypes & IEEE Frontiers in Education Conference, FIE 2020	\\
S19 & Schneider, J.-G., Eklund, P.W., Lee, K., Chen, F., Cain, A., Abdelrazek, M. & 2020 & Adopting industry agile practices in large-scale capstone education & 42nd International Conference on Software Engineering: Software Engineering Education and Training, ICSE-SEET 2020 \\
S20 & Ye, H. & 2009 & An academia-industry collaborative teaching and learning model for software engineering education	& 21st International Conference on Software Engineering and Knowledge Engineering, SEKE 2009	\\
S21 & Demuth, B., Kandler, M. & 2017 & An Approach for Project Task Approximation in a Large-Scale Software Project Course & 30th IEEE Conference on Software Engineering Education and Training, CSEE\&T 2017	\\
S22 & Ellis, H.J.C. & 2007 & An assessment of a self-directed learning approach in a graduate web application design and development course	& IEEE Transactions on Education \\
S23 & Anslow, C., Maurer, F. & 2015 & An experience report at teaching a group based agile software development project course & 46th ACM Technical Symposium on Computer Science Education \\
S24 & Bareiss, R., Katz, E. & 2011 & An exploration of knowledge and skills transfer from a formal software engineering curriculum to a capstone practicum project & 24th IEEE-CS Conference on Software Engineering Education and Training, CSEE\&T 2011 \\
S25 & Stephenson, B., James, M., Brooke, N., Aycock, J. & 2016 & An Industrial Partnership Game Development Capstone Course & 17th Annual Conference on Information Technology Education \\
\midrule
\bottomrule
\end{tabular}
\end{minipage}
\end{table}


\begin{table*}[width=1\textwidth,cols=5]
\scriptsize
\caption*{\textit{\small Continued from previous page}}
\begin{tabular}{p{0.07\linewidth}p{0.19\linewidth}p{0.03\linewidth}p{0.31\linewidth}p{0.28\linewidth}}
\toprule
ID	& Author(s) &	Year	& Title  & Source title \\
\midrule
S26 & Bell, J.T., Prabhu, A. & 2015 & An innovative approach to Software Engineering term projects, coordinating student efforts between multiple teams over multiple semesters & IEEE Frontiers in Education Conference, FIE 2014 \\
S27 & Vasilevskaya, M., Broman, D., Sandahl, K. & 2015 & Assessing large-project courses: Model, activities, and lessons learned & ACM Transactions on Computing Education, TOCE	\\
S28 & von Konsky, B.R., Ivins, J. & 2008 & Assessing the capability and maturity of capstone software engineering projects & Tenth conference on Australasian computing education - Volume 78	\\
S29 & Fontao, A., Gadelha, B., Junior, A.C. & 2019 & Balancing Theory and Practice in Software Engineering Education - A PBL, toolset based approach & IEEE Frontiers in Education Conference, FIE 2019	\\
S30 & Harding, T. & 2007 & Benefits and struggles of using large team projects in capstone courses & ASEE Annual Conference and Exposition	\\
S31 & Engelsma, J. R. & 2014 & Best practices for industry-sponsored CS capstone courses & Journal of Computing Sciences in Colleges \\ 
S32 & Matthies, C., Teusner, R., Hesse, G. & 2019 & Beyond Surveys: Analyzing Software Development Artifacts to Assess Teaching Efforts	 & IEEE Frontiers in Education Conference, FIE 2018	\\
S33 & Ziv, H., Patil, S. & 2010 & Capstone project: From software engineering to ``Informatics`` & 23rd IEEE Conference on Software Engineering Education and Training, CSEE\&T 2010 \\
S34 & Anderson, Ruth E.; Borriello, Gaetano; Martin, Hélène; Black, Leonard & 2009 & Capstone projects as community connectors & Journal of Computing Sciences in Colleges	\\
S35 & Paasivaara, M., Vanhanen, J., Lassenius, C. & 2019 & Collaborating with industrial customers in a capstone project course: The customers' perspective & IEEE/ACM 41st International Conference on Software Engineering: Software Engineering Education and Training, ICSE-SEET 2019	\\
S36 & Adams, R., Kleiner, C. & 2016 & Collaboration support in an international computer science capstone course & International Conference on Social Computing and Social Media \\
S37 & Watkins, K.Z., Barnes, T. & 2010 & Competitive and agile software engineering education & IEEE SoutheastCon, SoutheastCon 2010	\\
S38 & Gustavsson, H., Brohede, M. & 2019 & Continuous assessment in software engineering project course using publicly available data from GitHub & 15th International Symposium on Open Collaboration, OpenSym 2019	\\
S39 & Hadfield, Steven M.; Jensen, Nathan A. & 2007 & Crafting a software engineering capstone project course & Journal of Computing Sciences in Colleges \\
S40 & Rong, G., Shao, D. & 2012 & Delivering software process-specific project courses in tertiary education environment: Challenges and solution & 25th IEEE Conference on Software Engineering Education and Training, CSEE\&T 2012	\\
S41 & Nguyen, D.M., Truong, T.V., Le, N.B. & 2013 & Deployment of capstone projects in software engineering education at Duy Tan university as part of a university-wide project-based learning effort & Learning and Teaching in Computing and Engineering, LaTiCE 2013	\\
S42 & Lago, P., Schalken, J., Vliet, H.V. & 2009 & Designing a multi-disciplinary software engineering project & 22nd IEEE Conference on Software Engineering Education and Training, CSEE\&T 2009	\\
S43 & Angelov, S., de Beer, P. & 2017 & Designing and applying an approach to software architecting in agile projects in education & Journal of Systems and Software	\\
S44 & Anderson, R.E., Kolko, B. & 2011 & Designing technology for resource-constrained environments: A multidisciplinary capstone sequence & Frontiers in Education, FIE 2012	\\
S45 & Leilde, V., Ribaud, V. & 2017 & Does Process Assessment Drive Process Learning? the Case of a Bachelor Capstone Project & 30th IEEE Conference on Software Engineering Education and Training, CSEE\&T 2017	\\
S46 & Brown, Q., Lee, F., Alejandre, S. & 2009 & Emphasizing soft skills and team development in an educational digital game design course & 4th International Conference on the Foundations of Digital Games, FDG 2009	\\
S47 & Takala, T. M., Malmi, L., Pugliese, R., Takala, T. & 2016 & Empowering students to create better virtual reality applications: A longitudinal study of a VR capstone course & Informatics in Education \\
S48 & Marques, M., Ochoa, S.F., Bastarrica, M.C., Gutierrez, F.J. & 2018 & Enhancing the Student Learning Experience in Software Engineering Project Courses & IEEE Transactions on Education \\
S49 & De Souza, R.T., Zorzo, S.D., Da Silva, D.A. & 2015 & Evaluating capstone project through flexible and collaborative use of Scrum framework & Frontiers in Education Conference, FIE 2015 \\
S50 & Vu, J.H., Frojd, N., Shenkel-Therolf, C., Janzen, D.S. & 2009 & Evaluating test-driven development in an industry-sponsored capstone project & 6th International Conference on Information Technology: New Generations, ITNG 2009	\\
S51 & Laplante, P.A., Defranco, J.F., Guimaraes, E. & 2019 & Evolution of a graduate software engineering capstone course - A course review	 & International Journal of Engineering Education	\\
S52 & Lederman, Timoth C. & 2010 & Evolution of capstone-courses in software engineering a finishing school	& Journal of Computing Sciences in Colleges	\\
S53 & Delgado, D., Velasco, A., Aponte, J., Marcus, A. & 2017 & Evolving a Project-Based Software Engineering Course: A Case Study & 30th IEEE Conference on Software Engineering Education and Training, CSEE\&T 2017	\\
\midrule
\bottomrule
\end{tabular}
\end{table*}

\begin{table*}[width=1\textwidth,cols=5]
\scriptsize
\caption*{\textit{\small Continued from previous page}}
\begin{tabular}{p{0.07\linewidth}p{0.19\linewidth}p{0.03\linewidth}p{0.31\linewidth}p{0.28\linewidth}}
\toprule
ID	& Author(s) &	Year	& Title   & Source title \\
\midrule
S55 & Ras, Eric and Carbon, Ralf and Decker, Björn and Rech, Jörg & 2007 & Experience Management Wikis for Reflective Practice in Software Capstone Projects & IEEE Transactions on Education	\\
S56 & Schorr, R. & 2020 & Experience Report on Key Success Factors for Promoting Students' Engagement in Software Development Group Projects & 4th IEEE World Conference on Engineering Education, EDUNINE 2020 \\
S57 & Longstreet, C. Shaun; Cooper, Kendra & 2013 & Experience report: A sustainable serious educational game capstone project & CGAMES'2013 USA	\\
S58 & Dupuis, R., Champagne, R., April, A., Séguin, N. & 2010 & Experiments with Adding to the Experience that Can be Acquired from Software Courses & 7th International Conference on the Quality of Information and Communications Technology, QUATIC 2010	\\
S59 & Burge, J. & 2007 & Exploiting Multiplicity to Teach Reliability and Maintainability in a Capstone Project & 20th IEEE Conference on Software Engineering Education and Training, CSEE\&T 2007	\\
S60 & Marshall, L., Pieterse, V., Thompson, L., Venter, D.M. & 2016 & Exploration of Participation in Student Software Engineering Teams & ACM Transactions on Computing Education, TOCE \\
S61 & Ganci, A., Ramnath, R., Ribeiro, B., Stone, R.B. & 2011 & Exploring collaboration between computer science engineers and visual communication designers in educational settings & 13th International Conference on Engineering and Product Design Education, E\&PDE 2011	\\
S62 & Burden, H., Steghöfer, J.-P., Hagvall Svensson, O. & 2019 & Facilitating entrepreneurial experiences through a software engineering project course & 41st International Conference on Software Engineering: Software Engineering Education and Training, ICSE-SEET 2019	\\
S63 & Basholli, A., Baxhaku, F., Dranidis, D., Hatziapostolou, T. & 2013 & Fair assessment in software engineering capstone projects & 6th Balkan Conference in Informatics	\\
S64 & Magana, A. J., Seah, Y. Y., Thomas, P. & 2018 & Fostering cooperative learning with Scrum in a semi-capstone systems analysis and design course & Journal of Information Systems Education \\
S65 & Sievi-Korte, O., Systä, K., Hjelsvold, R. & 2015 & Global vs. local -- Experiences from a distributed software project course using agile methodologies & Frontiers in Education, FIE 2015	\\
S66 & Hebig, R., Ho-Quang, T., Jolak, R., Schröder, J., Linero, H., Ågren, M., Maro, S.H. & 2020 & How do students experience and judge software comprehension techniques? & 28th International Conference on Program Comprehension \\
S67 & Verdicchio, Michael & 2021 & Hurricanes and pandemics: an experience report on adapting software engineering courses to ensure continuity of instruction	& Journal of Computing Sciences in Colleges	\\
S68 & Włodarski, R., Poniszewska-Marańda, A., Falleri, J.-R. & 2022 & Impact of software development processes on the outcomes of student computing projects: A tale of two universities & Information and Software Technology	\\
S69 & Izu, Cruz & 2018 & Improving Outcomes for a Masters Capstone IT Project & IEEE International Conference on Teaching, Assessment, and Learning for Engineering, TALE 2018 \\
S70 & Flowers, J.G. & 2008 & Improving the Capstone project experience: a case study in software engineering & 46th Annual Southeast Regional Conference on XX	\\
S71 & Gannod, Gerald C.; Bachman, Kristen M.; Troy, Douglas A.; Brockman, Steve D. & 2010 & Increasing alumni engagement through the capstone experience & Frontiers in Education, FIE 2010	\\
S72 & Zilora, S.J. & 2015 & Industry-emulated projects in the classroom	& 16th Annual ACM Conference on Information Technology Education, SIGITE 2015	\\
S73 & Spichkova, M. & 2019 & Industry-oriented project-based learning of software engineering & 24th International Conference on Engineering of Complex Computer Systems, ICECCS 2019 \\
S74 & Carvalho, J.A., Sousa, R.D., Sá, J.O. & 2010 & Information systems development course: Integrating business, IT and IS competencies & 2010 IEEE Transforming Engineering Education: Creating Interdisciplinary Skills for Complex Global Environments	\\
S75 & Palacin-Silva, M.V., Seffah, A., Porras, J. & 2018 & Infusing sustainability into software engineering education: Lessons learned from capstone projects & Journal of Cleaner Production	\\
S76 & Kumar, S., Wallace, C. & 2015 & Instruction in software project communication through guided inquiry and reflection & Frontiers in Education, FIE 2015	\\
S77 & Zeid, A. & 2012 & Integrating international students' contests with computer science capstone: Lessons learned and best practices & Frontiers in Education, FIE 2012	\\
S78 & Lundqvist, K., Ahmed, A., Fridman, D., Bernard, J.-G. & 2019 & Interdisciplinary Agile Teaching & Frontiers in Education, FIE 2019	\\
S79 & Santoso, H.B., Lawanto, O., Purwandari, B., Isal, R.Y.K., Fitriansyah, R. & 2018 & Investigating Students' Metacognitive Skills while Working on Information Systems Development Projects & 7th World Engineering Education Forum, WEEF 2017	\\
S80 & Christensen, E.L., Paasivaara, M. & 2022 & Learning Soft Skills through Distributed Software Development & International Conference on Software and System Processes and Internation Conference on Global Software Engineering \\
S81 & Rout, Terence P.; Seagrott, John & 2007 & Maintaining High Process Capability in a Student Project Course & 20th Conference on Software Engineering Education \& Training, CSEE\&T 2007 \\
S82 & Rodriguez, G., Soria, A., Campo, M. & 2016 & Measuring the Impact of Agile Coaching on Students' Performance & IEEE Transactions on Education	\\
S83 & Linhoff, J., Settle, A. & 2009 & Motivating and evaluating game development capstone projects & 4th International Conference on Foundations of Digital Games \\
\midrule
\bottomrule
\end{tabular}
\end{table*}


\begin{table*}[width=1\textwidth,cols=5]
\scriptsize
\caption*{\textit{\small Continued from previous page}}
\begin{tabular}{p{0.07\linewidth}p{0.19\linewidth}p{0.03\linewidth}p{0.31\linewidth}p{0.28\linewidth}}
\toprule
ID	& Author(s)  &	Year	& Title   & Source title \\
\midrule
S84 & Haddad, H.M. & 2013 & One-semester CS capstone: A 40-60 teaching approach	& 10th International Conference on Information Technology: New Generations, ITNG 2013	\\
S85 & Fan, Xiaocong & 2018 & Orchestrating Agile Sprint Reviews in Undergraduate Capstone Projects & Frontiers in Education, FIE 2018 \\
S86 & Fagerholm, F., Vihavainen, A. & 2013 & Peer assessment in experiential learning: Assessing tacit and explicit skills in agile software engineering capstone projects & Frontiers in Education, FIE 2013 \\

S87 & Vasankari, T., Majanoja, A.-M. & 2019 & Practical Software Engineering Capstone Course – Framework for Large, Open-Ended Projects to Graduate Student Teams & Internation Conference on Computer Supported Education \\
S88 & Karunasekera, S., Bedse, K. & 2007 & Preparing software engineering graduates for an industry career & 20th Conference on Software Engineering Education \& Training, CSEE\&T 2007	\\
S89 & Weerawarana, S.M., Perera, A.S., Nanayakkara, V. & 2012 & Promoting creativity, innovation and engineering excellence: A case study from Sri Lanka & IEEE International Conference on Teaching, Assessment, and Learning for Engineering, TALE 2012	\\
S90 & Fornaro, R.J., Heil, M.R., Tharp, A.L. & 2007 & Reflections on 10 years of sponsored senior design projects: Students win-clients win! & Journal of Systems and Software	\\
S91 & Roach, S. & 2011 & Retrospectives in a software engineering project course: Getting students to get the most from a project experience & 24th IEEE-CS Conference on Software Engineering Education and Training, CSEE\&T 2011	\\
S92 & Mäkiaho, P., Poranen, T. & 2018 & Risks management in software development capstone projects & 19th International Conference on Computer Systems and Technologies \\
S93(a,b) & MacKellar, B. K., Sabin, M., Tucker, A. & 2013 & Scaling a framework for client-driven open source software projects: A report from three schools & Journal of Computing Sciences in Colleges \\
S94 & Yuen, T.T. & 2015 & Scrumming with educators: Cross-departmental collaboration for a summer software engineering capstone & International Conference on Learning and Teaching in Computing and Engineering, LaTiCE 2015	\\
S95 & Isomöttönen, V., Daniels, M., Cajander, Å., Pears, A., McDermott, R. & 2019 & Searching for global employability: Can students capitalize on enabling learning environments? & ACM Transactions on Computing Education	\\
S96 & Maxim, B. & 2008 & Serious games as software engineering capstone projects & ASEE Annual Conference and Exposition \\
S97 & Krogstie, B.R., Divitini, M. & 2009 & Shared timeline and individual experience: Supporting retrospective reflection in student software engineering teams & 22nd Conference on Software Engineering Education and Training, CSEE\&T 2009	\\
S98 & Johns-Boast, L., Flint, S. & 2013 & Simulating industry: An innovative software engineering capstone design course & Frontiers in Education, FIE 2013	\\
S99 & Boti, E., Damasiotis, V., Fitsilis, P. & 2021 & Skills Development Through Agile Capstone Projects & International Conference on Frontiers in Software Engineering \\
S100 & Paiva, S.C., Carvalho, D.B.F. & 2018 & Software creation workshop: A capstone course for business-oriented software engineering teaching & XXXII Brazilian Symposium on Software Engineering \\
S101 & Saeedi, K., Visvizi, A. & 2021 & Software development methodologies, HEIs, and the digital economy & Education Sciences \\
S102 & Smith, T., Cooper, K.M.L., Longstreet, C.S. & 2011 & Software engineering senior design course: Experiences with agile game development in a capstone project & International Conference on Software Engineering	\\
S103 & Jaccheri, L., Sindre, G. & 2007 & Software engineering students meet interdisciplinary project work and art & 11th International Conference on Information Visualisation, IV 2007	\\
S104 & Krusche, S., Dzvonyar, D., Xu, H., Bruegge, B. & 2018 & Software Theater—Teaching Demo-Oriented Prototyping & ACM Transactions on Computing Education, TOCE \\
S105 & Budd, A.J., Ellis, H.J.C. & 2008 & Spanning the gap between software engineering instructor and student & Frontiers in Education, FIE 2008	\\
S106 & Decker, A., Egert, C.A., Phelps, A. & 2016 & Splat! er, shmup? A postmortem on a capstone production experience & Frontiers in Education, FIE 2008 \\
S107 & Kerbs, R. & 2007 & Student teamwork: A capstone course in game programming & Frontiers in Education, FIE 2007	\\
S108 & Tadros, Ibrahem; Hammami, Samir; Al-Zoubi, Khaled & 2008 & Systems Development Projects & 3rd International Conference on Information and Communication Technologies: From Theory to Applications	\\
S109 & Jarzabek, S. & 2013 & Teaching advanced software design in team-based project course & 26th IEEE International Conference on Software Engineering Education and Training, CSEE\&T 2013 \\
S110 & Lu, Baochuan; DeClue, Tim & 2011 & Teaching agile methodology in a software engineering capstone course & Journal of Computing Sciences in Colleges	\\
S111 & Cagiltay, N.E. & 2007 & Teaching software engineering by means of computer-game development: Challenges and opportunities & British Journal of Educational Technology	\\
\midrule
\bottomrule
\end{tabular}
\end{table*}


\begin{table*}[width=1\textwidth,cols=5]
\scriptsize
\caption*{\textit{\small Continued from previous page}}
\begin{tabular}{p{0.07\linewidth}p{0.19\linewidth}p{0.03\linewidth}p{0.31\linewidth}p{0.28\linewidth}}
\toprule
ID	& Author(s) &	Year	& Title  & Source title \\
\midrule
S112 & Tafliovich, A., Caswell, T., Estrada, F. & 2019 & Teaching software engineering with free open source software development: An experience report & Annual Hawaii International Conference on System Sciences	\\
S113 & Paasivaara, M., Lassenius, C., Damian, D., Raty, P., Schroter, A. & 2013 & Teaching students global software engineering skills using distributed Scrum & 35th International Conference on Software Engineering, ICSE 2013 \\
S114 & Khmelevsky, Y. & 2016 & Ten years of capstone projects at Okanagan College: A retrospective analysis & 21st Western Canadian Conference on Computing Education \\
S115 & Mahnič, V. & 2015 & The capstone course as a means for teaching agile software development through project-based learning & World Transactions on Engineering and Technology Education	\\
S116 & Broman, D., Sandahl, K., Baker, M.A. & 2012 & The company approach to software engineering project courses & IEEE Transactions on Education	\\
S117 & Khakurel, J., Porras, J. & 2020 & The Effect of Real-World Capstone Project in an Acquisition of Soft Skills among Software Engineering Students & 32nd IEEE Conference on Software Engineering Education and Training, CSEE\&T 2020 \\
S118 & Iacob, C., Faily, S. & 2020 & The impact of undergraduate mentorship on student satisfaction and engagement, teamwork performance, and team dysfunction in a software engineering group project & 51st ACM Technical Symposium on Computer Science Education, SIGCSE 2020 \\
S119 & Hoar, R. & 2014 & The real world web: How institutional IT affects the delivery of a capstone web development course & 19th Western Canadian Conference on Computing Education, WCCCE 2014	\\
S120 & Yue, K. B., Damania, Z., Nilekani, R., Abeysekera, K. & 2011 & The use of free and open source software in real-world capstone projects & Journal of Computing Sciences in Colleges \\
S121 & Isomöttönen, V., Kärkkäinen, T. & 2008 & The value of a real customer in a capstone project & 21st Conference on Software Engineering Education and Training, CSEE\&T 2008 \\
S122 & Mohan, S., Chenoweth, S., Bohner, S. & 2012 & Towards a better capstone experience & 43rd ACM Technical Symposium on Computer Science Education, SIGCSE'12 \\
S123 & Rico, D.F., Sayani, H.H. & 2009 & Use of agile methods in software engineering education & Agile Conference, AGILE 2009	\\
S124 & Tribelhorn, B., Nuxoll, A.M. & 2021 & Using Agile and Active Learning in Software Development Curriculum & ASEE Virtual Annual Conference and Exposition	\\
S125 & McDonald, J., Wolfe, R. & 2008 & Using computer graphics to foster interdisciplinary collaboration in capstone courses & Journal of Computing Sciences in Colleges \\
S126 & Ju, A., Hemani, A., Dimitriadis, Y., Fox, A. & 2020 & What agile processes should we use in software engineering course projects? & 51st ACM Technical Symposium on Computer Science Education, SIGCSE 2020 \\
S127 & Bastarrica, M.C., Perovich, D., Samary, M.M. & 2017 & What can students get from a software engineering capstone course? & 39th IEEE/ACM International Conference on Software Engineering: Software Engineering and Education Track, ICSE-SEET 2017	\\
\midrule
\bottomrule
\end{tabular}
\end{table*}

\end{document}